\pdfoutput=1
\PassOptionsToPackage{hyphens}{url}
\documentclass[camera,letterpaper,nomarginnotes,nonarrowgutter]{jpaper}

\usepackage{fancyhdr}
\usepackage[numbers,sort&compress]{natbib}
\usepackage[colorlinks=true,urlcolor=blue,citecolor=blue,linkcolor=blue]{hyperref}
\usepackage{colortbl}          %
\usepackage{tabularx}
\usepackage{multirow}
\usepackage{array}
\usepackage{cleveref}          %
\usepackage[ruled,vlined,linesnumbered]{algorithm2e}
\usepackage{tikz}
\usetikzlibrary{arrows.meta,positioning,shapes.geometric,fit,calc}
\usepackage{enumitem}
\usepackage{subcaption}
\usepackage{textcomp}
\usepackage{pifont}
\usepackage{listings}
\usepackage{amsmath}
\usepackage{microtype}
\usepackage[textsize=scriptsize]{todonotes}
\usepackage{stfloats} %
\DeclareRobustCommand{\circled}[1]{\tikz[baseline=(char.base)]{\node[shape=circle,fill=white,draw=black,text=black,line width=0.5pt,inner sep=0.5pt] (char) {\small\textbf{#1}};}}
\DeclareRobustCommand{\blackcircled}[1]{\tikz[baseline=(char.base)]{\node[shape=circle,fill=black,draw=black,text=white,line width=0.5pt,inner sep=0.5pt] (char) {\small\textbf{#1}};}}
\DeclareRobustCommand{\redcircled}[1]{\tikz[baseline=(char.base)]{\node[shape=circle,fill=red!75!black,draw=red!75!black,text=white,line width=0.5pt,inner sep=0.5pt] (char) {\small\textbf{#1}};}}

\newcommand{\chiron}{\textsc{Trail}}
\newcommand{\virtuoso}{Virtuoso}
\newcommand{\sniper}{Sniper}

\definecolor{tblheader}{RGB}{231,238,247}
\definecolor{tblstripe}{RGB}{247,249,252}
\definecolor{tblhighlight}{RGB}{231,245,236}
\definecolor{takeawaybg}{RGB}{244,248,252}
\definecolor{takeawayborder}{RGB}{141,161,186}
\definecolor{revCommon}{RGB}{111,66,193}
\definecolor{revA}{RGB}{192,57,43}
\definecolor{revB}{RGB}{41,128,185}
\definecolor{revC}{RGB}{39,130,86}
\definecolor{revD}{RGB}{211,84,0}

\definecolor{changeblue}{RGB}{0,90,200}

\newif\ifshowchanges
\showchangesfalse
\definecolor{hsorange}{RGB}{230,120,0}
\ifshowchanges
  
\else
  
\fi
\ifshowchanges
  \makeatletter
  \newcommand{\advnote}[1]{{\let\marginpar\jp@marginpar\todo[color=red!30,linecolor=red,bordercolor=red,size=\tiny]{#1}}}
  \makeatother
\else
  \newcommand{\advnote}[1]{}
\fi

\newcommand{\takeawaybox}[2][]{%
  \vspace{3pt}\noindent%
  \setlength{\fboxsep}{6pt}%
  \fcolorbox{takeawayborder}{takeawaybg}{\parbox{0.95\columnwidth}{%
    \ifx\relax#1\relax\textbf{Takeaway.}\else\textbf{Takeaway \##1:}\fi{} #2}}%
  \vspace{5pt}%
}
\newcommand{\todorevnotes}[3][revCommon]{}
\newcommand{\revqnote}[2][revCommon]{}

\newcommand\VMcharacterization{~\cite{vm2,karakostas_characterIISWC,
isca2010-barr-trancache,5-levelpaging,contiguitas2023,radiantISMM21,bhattacharjeePACT2009,devirtualizingASPLOS2018,hash_dont_cache,virtualizationimplication,vm29,
vmhpcIISWC2024nick}}

\newcommand\VMlargepages{~\cite{park2020perforated,guvenilir2020tailored,ingensOSDI2016,talluriISCA1992,panwar2018making,panwar2019hawkeye,tridentMICRO2021,pham2015,mosaic2017MICRO,promotionHPCA2001,shadowpageISCA1998,duHPCA2015,vm42,vm43,partialMICRO2020,ganapathy98}}

\newcommand\VMcontiguity{~\cite{translationranger2019,karakostas2015,chloe2020,hybridtlbISCA2017,flexpointerTACO2023,contiguitas2023,vm6,vm2,dmtASPLOS2024}}

\newcommand\TieredMemory{~\cite{tpp,pond,hemem,memtis,tmts,Yan2019,het7,Lagar2019,datatiering-dulloor-eurosys16,radiantISMM21,yoon2012,banshee,meza2012enabling,ramos11,li2017utility,kokolisPageSeerHPCA2019,het2,lee2009,alloy_cache,wangPantheraPLDI2019,chenHolisticTOCS2021,olsonHetMemTACO2022,oliveiraOptaneAccess2023,nadigHarmonia2025,songPCMISMM2020,wenMobileHybridTCAD2020,zhao2014firm,salkhordeh2019analytical,janus,daxvm}}

\newcommand\DisaggregatedMemory{~\cite{Gao2016,Shan2018,Korolija2021,Wang2020,Zuo2021,Maruf2020,Lim2009,Zhang2020,Angel2020,Lim2012,Peng2020,Bindschaedler2020,Katrinis2016,Aguilera2017,Aguilera2018,Rao2016,Calciu2021,Adya2019,Pinto2020,Gu2017,Buragohain2017,Zervas2018}}

\newcommand\DataPrefetchSpatial{~\cite{jouppi_prefetch,baer2,stride,chen1995tc,ampm,footprint,sms,sms_mod,spp,vldp,bop,bingo,dspatch,mlop,ipcp,kpc,pmp,pathfinder,navarro2022berti}}
\newcommand\DataPrefetchTemporal{~\cite{markov,ghb,stems,isb,misb,domino,triage,dol,cooksey2002stateless,wenisch2010making,wenisch2005temporal,chilimbi2002dynamic,chou2007low,ferdman2007last,hu2003tcp,bekerman1999correlated,karlsson2000prefetching}}
\newcommand\DataPrefetchLearned{~\cite{peled_rl,hashemi2018learning,voyager,pythia}}
\newcommand\DataPrefetchMgmt{~\cite{fdp,classifying,sandbox,clip,ppf,ebrahimi2009techniques,ebrahimi_paware,lee2008prefetch,memory_bank,panda2016expert,seq_stream}}

\fancypagestyle{trailversion}{%
  \fancyhf{}%
  \fancyfoot[C]{\thepage}%
}

\begin{document}

\title{{\chiron{}: Scalable and Low-Cost Temporal TLB Prefetching \newline via Page-Table-Embedded Deltas}}

\author{
    Konstantinos Kanellopoulos\textsuperscript{1} \hspace{0.5em}
    Konstantinos Sgouras\textsuperscript{1} \hspace{0.5em}
    Harsh Songara\textsuperscript{1} \hspace{0.5em} 
    Rahul Bera\textsuperscript{1} \hspace{0.5em}
    Onur Mutlu\textsuperscript{1,2} \vspace{0.5em} \\
    \normalsize{\textsuperscript{1}ETH Z\"urich} \hspace{1em}
    \normalsize{\textsuperscript{2}New York University}
}

\maketitle

\def\aeavailable{}     %
\def\aefunctional{}    %
\def\aereproduced{}    %
\AddToHookNext{shipout/foreground}{%
  \put(\paperwidth-0.75in,-0.2in-12mm){\makebox[0pt][r]{%
    \href{https://www.acm.org/publications/policies/artifact-review-and-badging-current}{%
    \ifdefined\aeavailable\includegraphics[width=12mm]{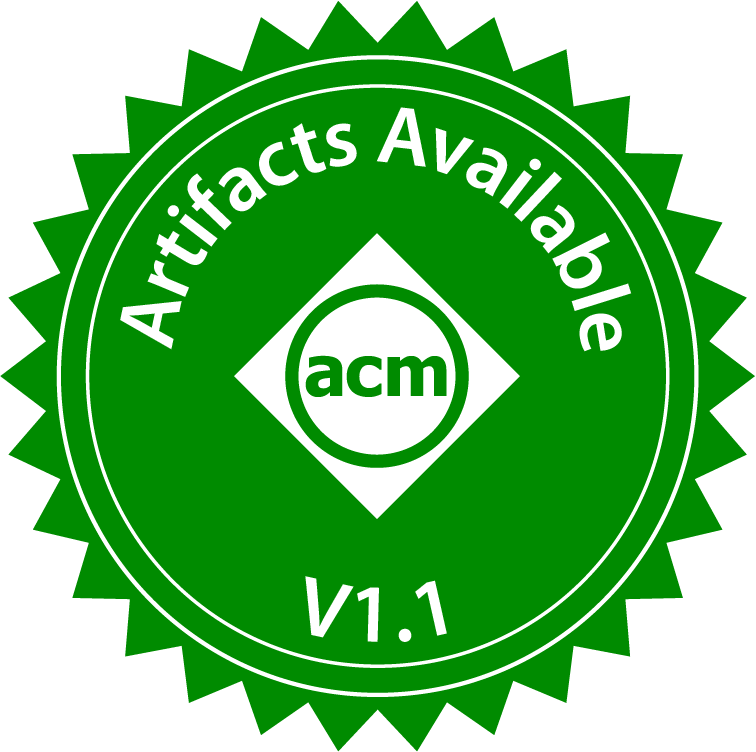}\fi
    \ifdefined\aefunctional\,\includegraphics[width=12mm]{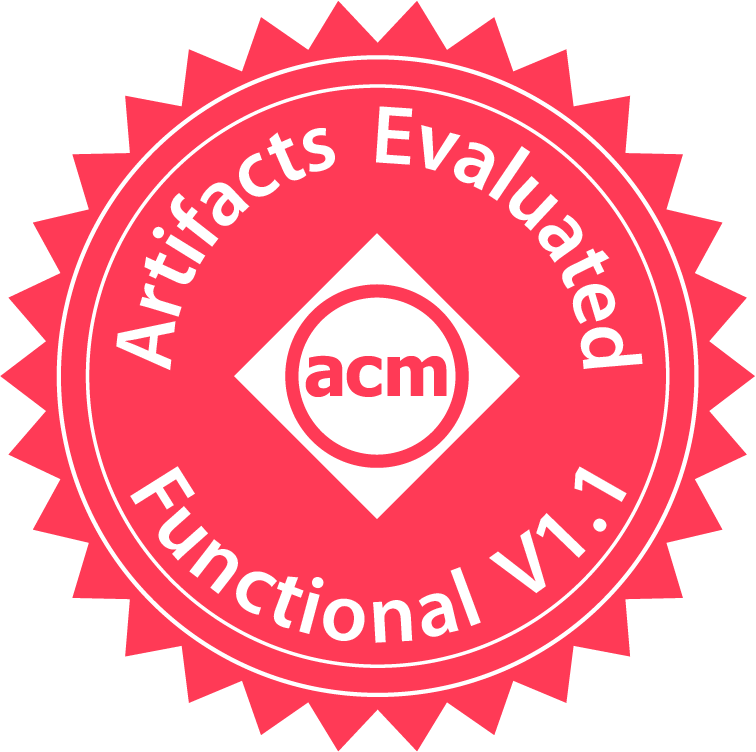}\fi
    \ifdefined\aereusable\,\includegraphics[width=12mm]{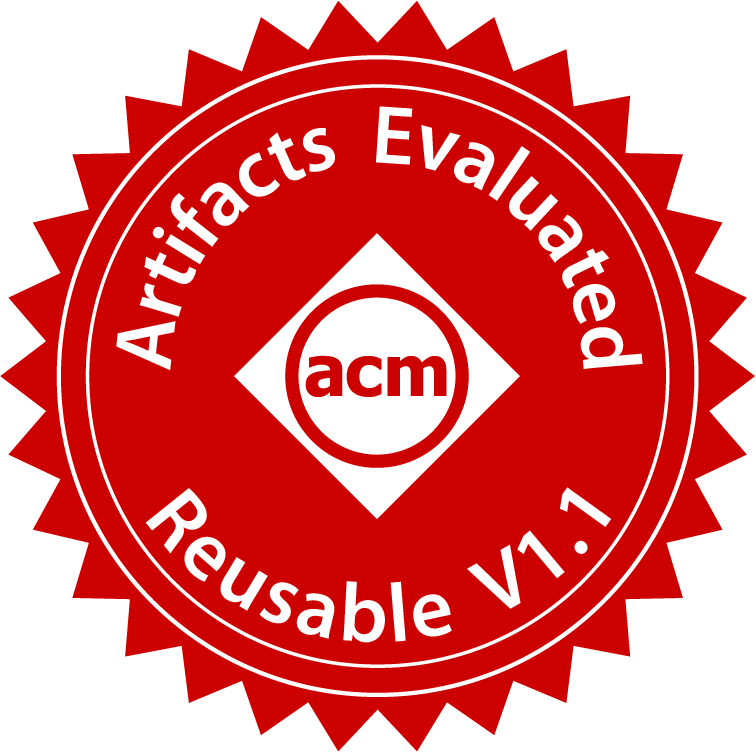}\fi
    \ifdefined\aereproduced\,\includegraphics[width=12mm]{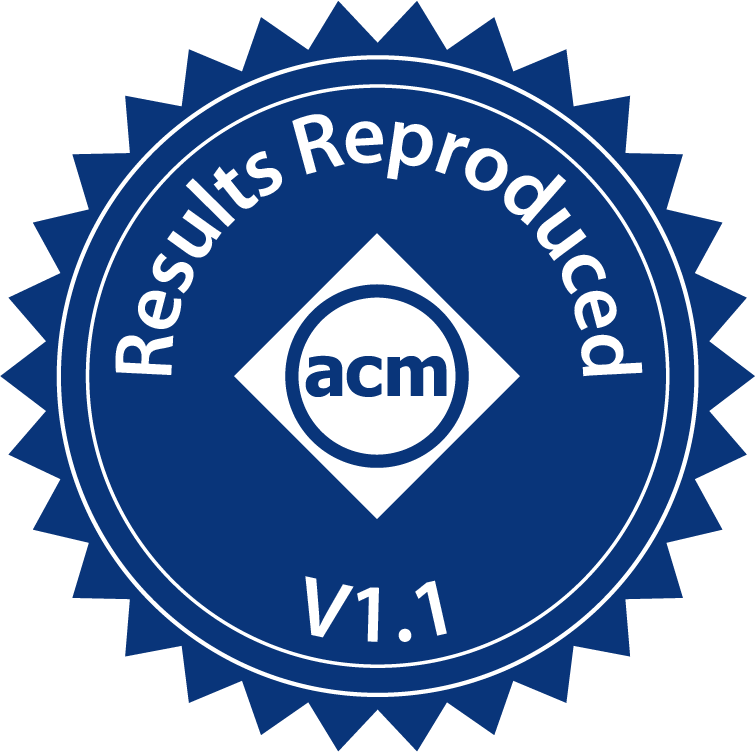}\fi}}}}%

\pagestyle{trailversion}
\thispagestyle{trailversion}

\begin{abstract}

Address translation is a significant performance bottleneck in modern data-intensive workloads.
Prefetching virtual-to-physical address mappings {into} the TLBs and the cache hierarchy
before they are demanded is a promising technique to hide address translation latency. 
However, accurately {capturing} address translation patterns is challenging as translation-intensive workloads often exhibit irregular memory access patterns.
Existing TLB prefetchers {have} two key shortcomings: (i) spatial prefetchers fail to capture irregular translation patterns, and (ii) temporal prefetchers record {address} deltas (i.e., distance{s} between virtual memory regions that successively trigger page table walks) 
in fixed-capacity hardware structures that cannot scale with the application's memory footprint.
In this work, we conduct a large-scale characterization study across 200 translation-intensive workloads and find that, despite the irregularity of the applications' memory accesses, deltas exhibit {a} highly predictable structure: (i) for a given virtual memory region, the set of candidate deltas is small and recurring, and 
(ii) more than 94\% of observed deltas fit in an 18-bit signed representation.

We introduce \chiron{}, a new temporal TLB prefetcher that embeds {deltas} directly in unused bits of the l{ast-level entries of the page table (PTEs)} to enable scalable, and low-cost prefetching.
When {a virtual memory region (the destination region) triggers a page table walk, \chiron{} identifies the region (the source region) that last triggered a walk from the same instruction and records the delta between the two in the source region's PTE.}
{The next time a page table walk fetches the source region's PTE cache block, 
\chiron{} retrieves the recorded deltas at no additional memory access cost, identifies the regions most likely to trigger the next page table walks, and prefetches the corresponding PTEs into the TLBs and the cache hierarchy.}
\chiron{} comes with three key benefits: 
 (i) {compactly storing multiple deltas in a single PTE cache block enables \chiron{} to issue
prefetches for multiple destination regions, improving prefetch coverage},
(ii) {embedding deltas in PTEs requires no additional storage in the processor or main memory}, and 
(iii) {since deltas reside in PTEs already allocated by the application, prefetch metadata capacity scales with the application's memory footprint at no extra cost}.

Our evaluation shows that \chiron{} improves performance in single-core (four-core) systems by 5.7\% (11.5\%) on average over the No-TLB-Prefetcher baseline across 200 translation-intensive workloads (100 multiprogrammed mixes), outperforming the best prior standalone TLB prefetcher by 1.7\% (2.5\%).
These gains require only a 64-entry hardware table in the processor to track the history of page table walks for each instruction. Trail is freely available at
\href{https://github.com/CMU-SAFARI/Virtuoso/tree/trail-artifact-release}{github.com/CMU-SAFARI/Virtuoso/TRAIL}.

\end{abstract}

\section{Introduction}
\label{sec:introduction}

As applications become more memory-intensive, address translation grows into a major performance bottleneck\VMcharacterization{}.
Recent studies from academia and industry demonstrate that address translation can account for up to 40--45\% of total execution time in modern data-intensive applications (15--20\% with large pages)~\cite{radiantISMM21,contiguitas2023,karakostas_characterIISWC,vm2,warehouse,dmtASPLOS2024,kanellopoulos2023utopia,kanellopoulosMICRO2023victima,elastic-cuckoo-asplos20,karakostas2015,chloe2020,vbi,midgard}.
We expect address translation overheads to grow even further as systems transition toward larger physical address spaces, driven by memory disaggregation\DisaggregatedMemory{} and tiered memory\TieredMemory{}.

\textbf{TLB Prefetching.} A promising technique to hide address translation latency is to predict which virtual-to-physical address mappings the application needs in the future,
pre-walk the page table to retrieve them
and insert the retrieved mappings into the TLBs and the cache hierarchy~\cite{Bala1994SoftwarePA,kandiraju2002going,vavouliotis2021,morriganMICRO2021,ha2025latpc,duong2025contiguity,baruah2020valkyrie,bhattacharjee2010intercore,chukhman2010contextAware,park1995softwareControlled}. 
We refer to a mechanism that performs such prefetching as a \emph{TLB prefetcher}.
The effectiveness of a TLB prefetcher depends on how accurately it captures the application's address translation patterns.
Doing so is challenging because translation-intensive workloads often exhibit irregular memory access and address translation patterns.

\textbf{Shortcomings of Prior Works.} Existing TLB prefetchers~\cite{kandiraju2002going,saulsbury2000recency,vavouliotis2021} fall short in two key ways. First, \emph{spatial} TLB prefetchers (e.g., next-page~\cite{Bala1994SoftwarePA}, IP-stride~\cite{chen1995tc}) rely on detecting spatial regularity and therefore fail to capture the \emph{irregular} translation patterns common in sparse and pointer-rich applications. Second, \emph{temporal} TLB prefetchers detect recurring patterns in translation requests or TLB evictions to predict future translation needs. Recency-based prefetching (REC)~\cite{saulsbury2000recency} links pages in L2 TLB eviction order using pointers stored in the page table and, on an L2 TLB miss, prefetches translations for pages adjacent to the accessed page (i.e., the one that missed in the L2 TLB) in this eviction order. However, pages evicted close in time are not necessarily accessed close in time in the future, making eviction order an unreliable basis for prefetching. Distance Prefetching (DP)~\cite{kandiraju2002going} uses a hardware table to record which delta tends to follow the current one, where a delta is the distance between pages that cause consecutive L2 TLB misses. However, this table has a fixed capacity that does not grow with the application's memory footprint. Translation-intensive workloads can exhibit more recurring delta pairs than the table can retain, causing DP to discard learned pairs before they recur and miss prefetching opportunities. This limits prefetch coverage (i.e., the fraction of TLB misses that the prefetcher successfully predicts). We discuss in detail both limitations in \S\ref{sec:bg:limitations}.

\textbf{Our goal} is to design a TLB prefetcher that (i)~effectively captures irregular address translation patterns, (ii)~has low hardware cost, and (iii)~scales prefetch coverage with the application's memory footprint.
Satisfying all three requirements concurrently is challenging.
Capturing irregular patterns across a large memory footprint requires the prefetcher to remember deltas for a large number of memory regions (requirements (i) and (iii)), while low hardware cost rules out the large dedicated tables that prior temporal prefetchers~\cite{kandiraju2002going,saulsbury2000recency} use to remember them (requirement (ii)).
Meeting all three requirements therefore hinges on a question that prior work leaves open: do the temporal translation patterns of an application have enough structure that a prefetcher can capture most of them by remembering only a few, compactly-encoded deltas per region?

\textbf{Large-Scale Characterization Study.} To answer this question, we conduct a large-scale characterization study in simulation across 200 translation-intensive workloads (\S\ref{sec:methodology} describes in detail our evaluation methodology).
We study the sequence of virtual addresses that trigger page table walks (PTWs).
Rather than tracking individual pages, we group virtual pages into fixed-size \emph{regions}.
A region is a contiguous chunk of virtual pages. Unless otherwise noted, we use 32\,KB regions, i.e., 8~contiguous 4\,KB pages.
We use four terms throughout the paper.
The \emph{source region} is the virtual memory region that contains the virtual address that triggered a PTW.
The \emph{destination region} is the virtual memory region that contains the virtual address of the next PTW in the sequence.
The \emph{delta} is the signed distance, in number of regions, from the source region to the destination region.
A \emph{transition} is a pair (source region, delta): it records that a PTW in the source region is followed by a PTW in the region at distance \emph{delta.}
We form transitions in two ways: \emph{global} tracking pairs each PTW with the immediately preceding PTW, whereas \emph{per-PC} tracking pairs it with the most recent preceding PTW triggered by the same instruction (identified by its program counter).

\textbf{Key Observations.} We make three key observations from our large-scale characterization study (described in detail in \S\ref{sec:bg:motivation}).
First, transitions are locally structured: a given source region has only a small, recurring set of destination regions, and this set is smallest when we track transitions per PC.
With per-PC tracking, the top-4 (top-8) most frequent destination regions of a source region cover ${\sim}$86\% (${\sim}$93\%) of all transitions.
A prefetcher therefore needs to remember only a few deltas per source region.
This structure reflects how irregular applications traverse and organize their data. An application's access pattern can be irregular while individual data-structure traversals exhibit recurring patterns. Irregular applications often traverse a few large data structures using a small set of loops. Within a loop, successive accesses by the same load instruction often target related elements stored in close proximity in virtual memory. For example, a graph traversal (e.g., Breadth-First Search) may visit vertices in an irregular order, yet scan each vertex's neighbors consecutively. Such locality concentrates PTW transitions from a source memory region (e.g., source vertex) into a small set of destination regions (e.g., destination vertices) (see the breadth-first-search example in \S\ref{sec:bg:motivation}).
Second, a larger memory region holds more pages (e.g., a 256\,KB region holds 64 pages, eight times more than a 32\,KB region), so a single delta covers transitions to more destination pages, which increases prefetch coverage.
However, prefetching a larger destination region also fetches the PTEs of more pages (e.g., 64 PTEs spread over eight cache blocks, versus 8 PTEs in one cache block), some of which the application may never access.
Third, deltas are compactly encodable: an 18-bit signed delta represents more than 94\% of all observed per-PC transitions. A few deltas (i.e., 2--4) per source region thus fit in a few tens of bits.
Together, these observations answer our question positively: despite their apparent irregularity, temporal translation patterns have enough structure that a prefetcher captures most of them by remembering only 2--4 18-bit deltas per source region. 

\textbf{Key Idea.} Based on these observations, we propose \textbf{\chiron{}},\footnote{We name our mechanism \chiron{} because it leaves a trail of deltas in the page table, which later PTWs follow to find the translations the application needs next.} whose key idea is to store the small set of deltas for each virtual memory region directly in the unused bits of the region's last-level page table entries (PTEs).\footnote{A leaf PTE has 64 bits in both x86-64~\cite{intelx86manual} and ARMv8~\cite{arm2013}. In x86-64, bits 58:52 are architecturally ignored by the hardware and thus free for software use. In ARMv8, bits 58:55 of a level-3 page descriptor are reserved for software use. In addition, both architectures define physical-address fields that can address up to 4\,PB, while the largest commercial servers today have roughly 32\,TB of memory~\cite{gcloud_x4,azure_mdsv3}, so the physical-address bits above the platform's maximum physical address width are unused. \chiron{} relies on a hardware/software contract in which the OS reserves a subset of these bits for \chiron{} and the page table walker ignores them during address translation. The number of available bits depends on the platform and the OS.}
Such an approach has three benefits.
First, the space for storing deltas scales with the application's memory footprint.
For each region, \chiron{} stores only a small set of deltas (e.g., the four most frequent), each with a confidence counter.
We call this set the region's \emph{payload}. The OS already allocates leaf PTEs for every virtual memory region the application allocates. Thus, every source region already has PTEs in which to store its deltas, and the space for deltas grows with the memory footprint, whereas a fixed-capacity hardware table overflows once the application's regions outnumber the table's entries.
Second, \chiron{}'s hardware cost is low, since the deltas occupy bits that already exist but are unused, rather than requiring a dedicated hardware table or additional memory.
Third, \chiron{} enables retrieving the whole payload (i.e., multiple deltas) with low latency and no additional memory accesses, 
since the PTW that fetches a region's PTEs also fetches the region's payload.
Thus, \chiron{} can immediately use the retrieved payload to determine which destination regions to prefetch, ensuring high prefetch coverage and timely prefetches.
For example, in x86-64, the 8~leaf PTEs of a 32\,KB region share one 64-byte cache block, and \chiron{} stores a payload of four 18-bit deltas, each with a 2-bit confidence counter, in this cache block, and retrieves it when a PTW fetches the block.\footnote{\chiron{} does \emph{not} require a fixed number of unused PTE bits. When a platform exposes fewer bits, \chiron{} stores fewer or narrower deltas per region. When a platform exposes no bits, \chiron{} stores the same per-region deltas in a separate OS-managed table that mirrors the structure of the page table, at the cost of one additional lookup per PTW (\S\ref{sec:mech:encoding}).}

\begin{figure}[h!]
  \centering
  \begin{tikzpicture}[font=\scriptsize, >=Latex,
      ev/.style={circle, fill=black, text=white, font=\tiny\bfseries, inner sep=0.7pt},
      walk/.style={align=center, font=\scriptsize\bfseries},
      act/.style={align=center, text width=1.85cm, inner sep=1pt}]
    \draw[thick] (0,0) -- (3.65,0);
    \draw[thick] (3.65,-0.1) -- (3.8,0.1);
    \draw[thick] (3.8,-0.1) -- (3.95,0.1);
    \draw[thick,->] (3.95,0) -- (7.7,0) node[right] {time};
    \node[anchor=west, font=\scriptsize\itshape] at (0,0.75) {Page-table walks from the same PC};
    \node[ev] (e1) at (0.95,0) {1};
    \node[ev] (e2) at (2.85,0) {2};
    \node[ev] (e3) at (4.85,0) {3};
    \node[ev] (e4) at (6.75,0) {4};
    \node[walk, above=2pt of e1] {PTW for $A$};
    \node[walk, above=2pt of e2] {PTW for $B$};
    \node[walk, above=2pt of e3] {PTW for $A$};
    \node[walk, above=2pt of e4, text=green!45!black] {Access to $B$};
    \node[act, below=2pt of e1] (a1) {PC table\\records $A$};
    \node[act, below=2pt of e2] (a2) {Record $\textcolor{blue!70!black}{\Delta = B - A}$\\in $A$'s payload};
    \node[act, below=2pt of e3] (a3) {Read $\textcolor{blue!70!black}{\Delta}$ from\\$A$'s payload\\\vspace{1pt} and prefetch $B = A + \textcolor{blue!70!black}{\Delta}$};
    \node[act, below=2pt of e4, text=green!45!black] (a4) {\ding{51} Hit in L2 TLB (or prefetch buffer), no PTW};
    \draw[->, dashed, blue!70!black] (a2.south) to[out=-90, in=-90, looseness=0.8]
      node[below, pos=0.5, yshift=-1pt] {Payload used later} (a3.south);
  \end{tikzpicture}
  \caption{Timeline of \chiron{}'s operations to record deltas and prefetch page-table entries.}
  \label{fig:timeline}
\end{figure}
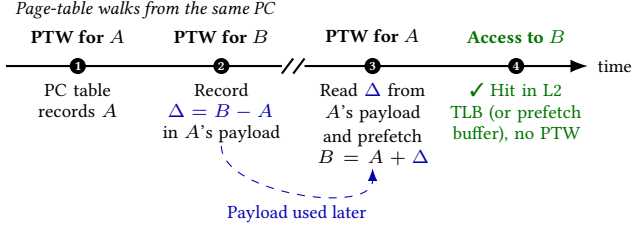

\Cref{fig:timeline} shows how \chiron{} works using an example sequence of PTWs triggered by the same PC.
On every PTW, \chiron{} performs two actions: it records a delta and it performs a prefetch (assuming confident).
To record a delta, \chiron{} needs to know the source region (i.e., the region that triggered the previous PTW from the same PC). 
Thus, \chiron{} maintains, as part of the core's memory management unit (MMU), a small 64-entry hardware \emph{PC table}  that stores, for each PC, the region that last triggered a PTW from that PC.
The PTW for $A$ (\blackcircled{1}) updates the PC table to record $A$ as the last region triggered by that PC.
The walk for $B$ (\blackcircled{2}) then discovers $A$ in the PC table, so $A$ is the source region and $B$ is the current region.
\chiron{} computes the delta $\Delta = B - A$ and records it in $A$'s payload.
A frequency-based replacement policy keeps the most frequent deltas in a payload and ages out stale ones.
To prefetch, \chiron{} reads the payload of the current region.
The payload arrives together with the region's PTEs, so reading it costs no additional memory access.
In the example, the second walk for $A$ (\blackcircled{3}) fetches $A$'s PTEs and with them $A$'s payload, which holds $\Delta$.
\chiron{} ranks the deltas in the payload by confidence, computes the corresponding destination regions, $B = A + \Delta$, and issues pre-walks to fetch their PTEs.
A pre-walk prefetches (i) the direct virtual-to-physical address mapping into
the L2 TLB (or in a prefetch buffer adjacent to the L2 TLB) and (ii) the corresponding PTEs into the cache hierarchy.
When the application next accesses $B$ (\blackcircled{4}), either the translation entry is already in the L2 TLB (or in the prefetch buffer) and no PTW is needed, or a demand PTW resolves faster since the required PTEs  already reside in the cache hierarchy.

\textbf{Evaluation Methodology.} We evaluate \chiron{} in single-core and multicore systems using Virtuoso~\cite{kanellopoulos2025virtuoso,virtuoso_github}, a validated simulation methodology that enables full-system simulation with high accuracy, ported on top of Sniper~\cite{sniper}, an event-driven multicore simulator.
We use 200 translation-intensive workloads from seven benchmark suites: CVP-1~\cite{cvp1} (Qualcomm traces), Graph analytics~\cite{gapbs,graphbig,ligra}, Google datacenter traces~\cite{googletraces}, XSBench~\cite{xsbench}, SPEC CPU~\cite{spec2017,spec2006}, Huawei DPC-4 ~\cite{dpc4}, and PARSEC Canneal~\cite{bienia2008parsec}.
We compare \chiron{} against six state-of-the-art TLB prefetchers (IP-Stride~\cite{chen1995tc}, Next-Page~\cite{Bala1994SoftwarePA}, DP~\cite{kandiraju2002going}, Berti~\cite{navarro2022berti}, Recency~\cite{saulsbury2000recency}, and ATP~\cite{vavouliotis2021}), a system with a larger L2 TLB, and a system with a perfect L2 TLB (i.e., one that always hits).

\textbf{Key Results.} Our evaluation yields four key results that showcase \chiron{}'s effectiveness.
First, in a single-core (four-core) system, \chiron{} improves performance by 5.7\% (11.5\%) on average over the No-TLB-Prefetcher baseline and outperforms the best prior standalone TLB prefetcher, Recency, by 1.7\% (2.5\%). Combining \chiron{} with Next-Page increases the improvement over No-TLB-Prefetcher to 7.6\% (14.1\%).
Second, across the 50 most translation-intensive workloads, \chiron{} reduces total demand PTW latency per kilo instructions by  60\%, compared with 42\% for the best prior prefetcher, Recency, since it both eliminates PTWs and significantly shortens the walks it does not eliminate.
Third, \chiron{}'s benefits are robust across different memory configurations and system parameters:
(i) \chiron{} improves performance by 4.4\% (1.6\%) on average compared to No-TLB-Prefetcher when 19\% (74\%) of memory is mapped with 2\,MB pages,
(ii) it improves performance across the entire evaluated DRAM bandwidth range (400--4800\,MTPS) compared to No-TLB-Prefetcher, and
(iii) it does not degrade performance on any of the 200 evaluated workloads.
Fourth, \chiron{} incurs very small hardware cost: it requires only a 64-entry PC table in the processor.

This paper makes the following contributions:
\begin{itemize}
    \item We conduct a large-scale experimental study of address translation patterns across 200 translation-intensive workloads. Our study shows that, despite their apparent irregularity, temporal translation patterns exhibit exploitable structure: each virtual memory region has a small, recurring, and compactly encodable set of per-PC deltas.

    \item We introduce \chiron{}, a temporal TLB prefetcher that stores each virtual memory region's deltas in unused bits of that region's PTEs. This design (i)~scales delta storage with the application's memory footprint, (ii)~requires neither dedicated SRAM nor additional memory capacity to store the deltas, and (iii)~retrieves the deltas together with the PTEs during a page table walk, \emph{without} additional memory accesses. This way, \chiron{} enables high prefetch coverage and timely prefetches with low hardware cost.

    \item We evaluate \chiron{} via simulation across 200 single-core workloads and 100 four-core multiprogrammed mixes against six state-of-the-art TLB prefetchers. Our results show that \chiron{} provides significant performance gains over the best prior TLB prefetcher in both single-core and multicore systems, while requiring only a 64-entry PC table. \chiron{} is freely available at \href{https://github.com/CMU-SAFARI/Virtuoso/tree/trail-artifact-release}{github.com/CMU-SAFARI/Virtuoso/TRAIL}.
\end{itemize}

\section{Background}
\label{sec:background}

\subsection{Address Translation Basics}
\label{sec:bg:translation}

A process accesses memory using virtual addresses (VAs)~\cite{denning1970,jacob1998}.
The OS gives each process its own VA space and maps it onto physical memory at the granularity of pages (e.g., 4\,KB)~\cite{denning1970,jacob1998}.
Address translation is the process of converting the VA of a memory access into the physical address (PA) that the memory hierarchy uses.
Every memory access requires address translation and thus address translation latency lies on the critical path of every memory access.
The OS records the VA-to-PA mapping of each page in a per-process \emph{page table} that resides in main memory~\cite{jacob1998}.
Because a VA space is large and sparsely used, the page table is organized as a radix tree, with four or five levels in x86-64~\cite{intelx86manual,5-levelpaging}.
Each non-leaf entry points to the next level of the tree, and each leaf \emph{page table entry} (PTE) stores the physical frame number and the protection bits of one virtual page.
Entries are 8 bytes~\cite{intelx86manual,arm2013}, so eight consecutive leaf PTEs, which map 32\,KB of contiguous virtual memory are stored in the same 64-byte cache block.
The OS allocates the pages of the page table on demand, so the page table grows with the memory footprint of the process.

The memory management unit (MMU) is the hardware component the CPU employs to perform address translation.
To avoid consulting the page table on every access, the MMU caches recently used mappings in a hierarchy of \emph{translation lookaside buffers} (TLBs)~\cite{denning1970,jacob1998}: small separate L1 TLBs for instructions and data, and a larger unified L2 TLB~\cite{intelx86manual}.
A memory access that misses in all TLBs triggers a \emph{page table walk} (PTW): a hardware circuit traverses the page table from the root to the leaf PTE, performing one memory access per level.
Page-walk caches~\cite{isca2010-barr-trancache} hold recently used non-leaf entries to reduce the number of memory accesses during a PTW, but only help when the non-leaf entries exhibit high reuse.

\subsection{TLB Prefetching Basics}
\label{sec:bg:tlb_prefetching}

A \emph{TLB prefetcher} predicts which VA-to-PA address mappings the application needs in the future, walks the page table in advance to retrieve them, and inserts the retrieved mappings into the TLBs and the cache hierarchy before the application demands them.
For each predicted page, the prefetcher issues a \emph{prefetch walk}, which traverses the page table like a demand walk but off the critical path.
The prefetcher inserts the resulting translation either directly into the L2 TLB (e.g.,~\cite{saulsbury2000recency}) or into a dedicated \emph{prefetch buffer} (e.g.,~\cite{kandiraju2002going,jouppi_prefetch}).
A prefetch walk has two benefits.
First, an L2 TLB miss that hits in the prefetch buffer needs no demand walk.
Second, it brings the page's leaf PTE cache block into the cache hierarchy, so even if the prefetch buffer has already evicted the translation, a later demand walk finds the PTE in the cache and completes faster.
Three metrics determine the effectiveness of a TLB prefetcher: (i) coverage, (ii) accuracy, and (iii) timeliness.
\emph{Coverage} is the fraction of demand walks that the prefetcher eliminates.
\emph{Accuracy} is the fraction of prefetched translations that the application uses (either via hits in the prefetch buffer or via hits in cached PTEs).
\emph{Timeliness} measures how early/late a prefetched translation arrives relative to when the application actually needs it.

\section{Motivation: Limitations of Existing TLB Prefetchers}
\label{sec:bg:limitations}

Prior TLB prefetchers fall into three categories: (i) spatial~\cite{Bala1994SoftwarePA,chen1995tc}, (ii) hybrid~\cite{vavouliotis2021}, and (iii) temporal~\cite{kandiraju2002going,saulsbury2000recency,morriganMICRO2021}.

\textbf{Spatial prefetchers} detect and exploit regular spatial address translation patterns to prefetch the virtual-to-physical mapping of pages
that are likely to be accessed next.
For example, a next-page prefetcher prefetches the virtual-to-physical mapping of the page that follows the current one.
A PC-stride (also called IP-stride) prefetcher~\cite{chen1995tc} learns, for each program counter (PC), the constant distance between successive pages accessed by that PC and uses this stride to predict the next page accesses and prefetch their translations.
However, many data-intensive applications with significant address translation overheads, such as graph analytics~\cite{ahnISCA2015,gapbs,graph500,graphbig,ainsworth2016graph,gnnSurvey,smash,besta2021sisa}, sparse machine learning~\cite{recommender,recommender1,dlmr,deeprec}, and other pointer-chasing workloads~\cite{db1,db2,db3,db4,mutlu2005address,roth1999effective,impicaICCD2016}, involve both regular and irregular memory access patterns, limiting the effectiveness of prefetchers that rely solely on spatial locality.

\textbf{Hybrid TLB prefetchers} combine complementary prefetchers and dynamically select among which prefetcher to use based on their recent effectiveness. For example, the \emph{Agile TLB Prefetcher (ATP)}~\cite{vavouliotis2021} combines (i)~a spatial prefetcher that fetches translations at fixed page offsets of $\pm1$ and $\pm2$, (ii)~a prefetcher that uses the two most recent deltas in the global L2 TLB miss stream, and (iii)~a PC-indexed stride prefetcher that uses the two most recent strides between L2 TLB misses caused by the same PC. ATP monitors candidate usefulness using shadow queues to select a component prefetcher for each L2 TLB miss, and disables prefetching when recent candidates are insufficiently useful.
However, ATP is only as effective as its constituent prefetchers. ATP's prefetchers rely on fixed spatial offsets or short delta histories, which fundamentally limits their ability to capture irregular translation patterns that involve multiple recurring deltas across a large number of virtual memory regions.

\textbf{Temporal prefetchers} detect recurring patterns in address translation requests~\cite{kandiraju2002going} or TLB evictions~\cite{saulsbury2000recency} to predict which virtual-to-physical mappings the application needs next.
Existing designs suffer from two key limitations. 
First, schemes like Recency-based prefetching (REC)~\cite{saulsbury2000recency}, rely on the order of L2 TLB evictions to predict future accesses, but L2 TLB eviction order does \emph{not} reflect the order in which the application will access pages in the future.
The eviction order is determined by the L2 TLB replacement policy, which selects which translation to evict based on multiple criteria (e.g., recency, frequency, and other heuristics) that are \emph{not} directly related to the application's actual access pattern.
Thus, pages whose translations are evicted close in time do not necessarily get accessed in close temporal proximity in the future. We evaluate REC extensively in \S\ref{sec:results}, in single-core and multi-core systems (\Cref{fig:prefetcher_comparison,fig:scurve,fig:coverage_decomposition,fig:multicore,fig:bw_sweep,fig:ablation}). In an extended version of the paper~\cite{trail_extended}, we analyze the limitations of REC in detail and show that (i) L2 TLB eviction order is not a reliable predictor of future page accesses and leads to inaccurate prefetching decisions, and (ii) REC requires expensive pointer-chasing structures to track recency information accurately.
Second, techniques like Distance Prefetching (DP)~\cite{kandiraju2002going}, heavily rely on small fixed-size on-chip hardware tables to store recurring delta (i.e., the distance between the virtual pages that cause consecutive L2 TLB misses) patterns. 
DP stores pairs of deltas that tend to follow one another in a fixed-size hardware table.
However, translation-intensive applications can exhibit more recurring pairs than DP's table can retain.
To make room for new pairs, DP may discard learned pairs before they recur, missing opportunities to prefetch.
We evaluate DP in \S\ref{sec:results} in single-core and multi-core systems (\Cref{fig:prefetcher_comparison,fig:coverage_decomposition,fig:multicore}).
In the extended version of this paper~\cite{trail_extended}, we further analyze the limitations of DP in detail and show that (i) a fixed-size hardware table cannot retain the history that translation-intensive workloads exhibit, no matter which program feature is used to index it (e.g. PC, virtual page number), and
(ii) a temporal TLB prefetcher requires history capacity that grows with the application’s memory footprint to
maintain high prediction accuracy.
We conclude that existing temporal prefetchers face fundamental limitations in capturing recurring address translation patterns.

\section{Goal, Opportunity and New Observations}
\label{sec:bg:motivation}

\textbf{Our goal} in this work is to design a TLB prefetcher that accurately captures recurring irregular translation patterns, incurs low hardware cost, and scales its history capacity with the application's memory footprint. 

Meeting this goal requires both a compact representation that preserves useful prediction information and a mechanism that retains this information \emph{without} requiring a large dedicated hardware table or additional storage in main memory.
To identify opportunities for storing \emph{temporal history} (i.e., the sequence of past page table walks) in a compact way, we conduct a large-scale characterization study of address translation patterns across 200 translation-intensive workloads in simulation. We examine whether address translation patterns exhibit enough structure to capture temporal correlations using only a small number of compact deltas (i.e., distances between memory regions that trigger consecutive page table walks) per memory region. In this section, we first define the terminology used throughout our study and then present our key observations.

\textbf{Terminology.}
We study the sequence of virtual addresses that trigger demand page table walks (PTWs).
We group virtual pages into fixed-size \emph{regions} of eight contiguous 4\,KB pages (32\,KB) unless otherwise noted.
For a pair of PTWs to regions $R_i$ and $R_j$, we call $R_i$ the \emph{source region}, $R_j$ the \emph{destination region}, and $\Delta=R_j-R_i$ the \emph{delta} and the tuple $(R_i,\Delta)$ represents the resulting \emph{transition}.
We form these pairs in two ways: \emph{global} tracking pairs each PTW with the immediately preceding PTW, whereas \emph{per-PC} tracking pairs it with the most recent preceding PTW triggered by the same PC.
The example below illustrates both schemes for three PTWs: PC A accesses region 4, PC B accesses region 9, and PC A accesses region 6.
Global tracking records transitions $(4,+5)$ and $(9,-3)$ between consecutive PTWs in the example.
Per-PC tracking instead records $(4,+2)$ between PC A's two PTWs and PC B produces no transition because it has no preceding PTW from the same PC.

\begingroup
\definecolor{seqA}{RGB}{35,89,125}
\definecolor{seqB}{RGB}{139,82,38}
\par\smallskip
\noindent\makebox[\columnwidth][c]{%
\begin{tikzpicture}[font=\scriptsize, >=Latex,
    event/.style={rectangle, rounded corners=2pt, line width=0.6pt,
                  minimum width=1.2cm, minimum height=0.72cm,
                  inner sep=3pt, align=center},
    eventA/.style={event, draw=seqA, fill=seqA!7},
    eventB/.style={event, draw=seqB, fill=seqB!7},
    delta/.style={above=2pt, font=\scriptsize, inner sep=1pt}]
  \node[anchor=east, align=right] at (0.95,0) {\textbf{Global}};
  \node[anchor=east, align=right] at (0.95,-1.25) {\textbf{Per-PC}\\(PC A)};
  \node[eventA] (globalSource) at (1.85,0)
    {\textcolor{seqA}{\textbf{PC A}}\\Region 4};
  \node[eventB] (globalMiddle) at (4.15,0)
    {\textcolor{seqB}{\textbf{PC B}}\\Region 9};
  \node[eventA] (globalDestination) at (6.45,0)
    {\textcolor{seqA}{\textbf{PC A}}\\Region 6};
  \node[eventA] (pcSource) at (1.85,-1.25)
    {\textcolor{seqA}{\textbf{PC A}}\\Region 4};
  \node[eventA] (pcDestination) at (6.45,-1.25)
    {\textcolor{seqA}{\textbf{PC A}}\\Region 6};
  \draw[densely dotted, black!35] (globalSource.south) -- (pcSource.north);
  \draw[densely dotted, black!35] (globalDestination.south) -- (pcDestination.north);
  \draw[->, semithick] (globalSource.east) -- node[delta] {$\Delta=+5$} (globalMiddle.west);
  \draw[->, semithick] (globalMiddle.east) -- node[delta] {$\Delta=-3$} (globalDestination.west);
  \draw[->, semithick, seqA] (pcSource.east) -- node[delta] {$\Delta=+2$} (pcDestination.west);
\end{tikzpicture}%
}
\par\smallskip
\endgroup

\textbf{Large-Scale Characterization Study.}
We make three key observations based on our \emph{large-scale characterization} of address translation patterns, performed in simulation using Virtuoso~\cite{kanellopoulos2025virtuoso,virtuoso_github}, across 200 translation-intensive workloads (see \S\ref{sec:methodology} for more detailed benchmarking and evaluation simulation methodology).

\begin{figure}[h!]
  \centering
  \includegraphics[width=1.0\columnwidth]{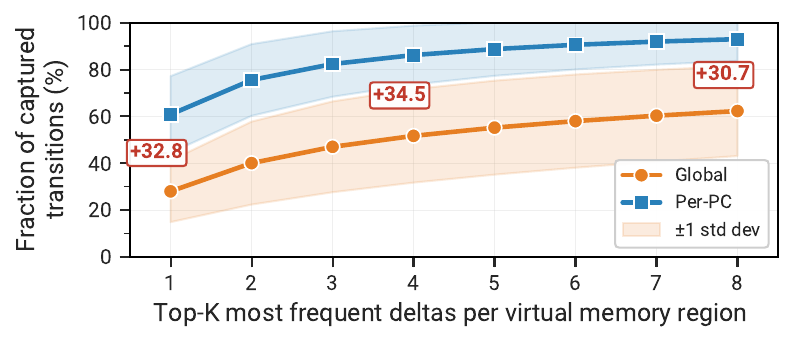}
  \caption{Average fraction of source-to-destination transitions captured by retaining the $k$ most frequent deltas per source region, with deltas computed by global or per-PC tracking, across 200 translation-intensive workloads.}
  \label{fig:motivation_pc_cond}
\end{figure}

\textbf{Observation~1: Storing 4-8 deltas per source memory region captures most source-to-destination transitions.}\label{obs:local_structure}\label{obs:compact_successors}
To test whether a small amount of history per source region suffices, we retain, for each 32\,KB source region, only the $k$ most frequent deltas observed from that region. We capture a pair of page table walks if the delta from the source region to the destination region is among the retained deltas.
We perform the experiment with deltas computed by global tracking and by per-PC tracking, as defined above.
\Cref{fig:motivation_pc_cond} shows the average fraction of captured transitions as $k$ grows from 1 to 8, across the 200 workloads.
We make two observations.
First, with per-PC tracking, a few deltas per source region capture most source-to-destination transitions: the four most frequent deltas capture 86\% of transitions, and the eight most frequent capture 93\%.
Second, per-PC tracking captures 31 to 36 percentage points more transitions than global tracking at every $k$.
The reason is that the PC identifies the code (e.g., a traversal loop) that produces a transition, so per-PC deltas connect regions that the same code accesses in sequence, whereas global tracking interleaves the PTWs of different PCs and thus computes deltas between potentially-unrelated regions.

\takeawaybox[1a]{A small number (4--8) of per-PC deltas for each source memory region captures most source-to-destination transitions in address translation patterns. A scalable prefetcher should therefore distribute delta storage across memory regions, allowing its history capacity to grow with the application's memory footprint rather than remain bounded by a fixed-size hardware table.}

\begin{figure}[h!]
  \centering
  \includegraphics[width=1.0\linewidth]{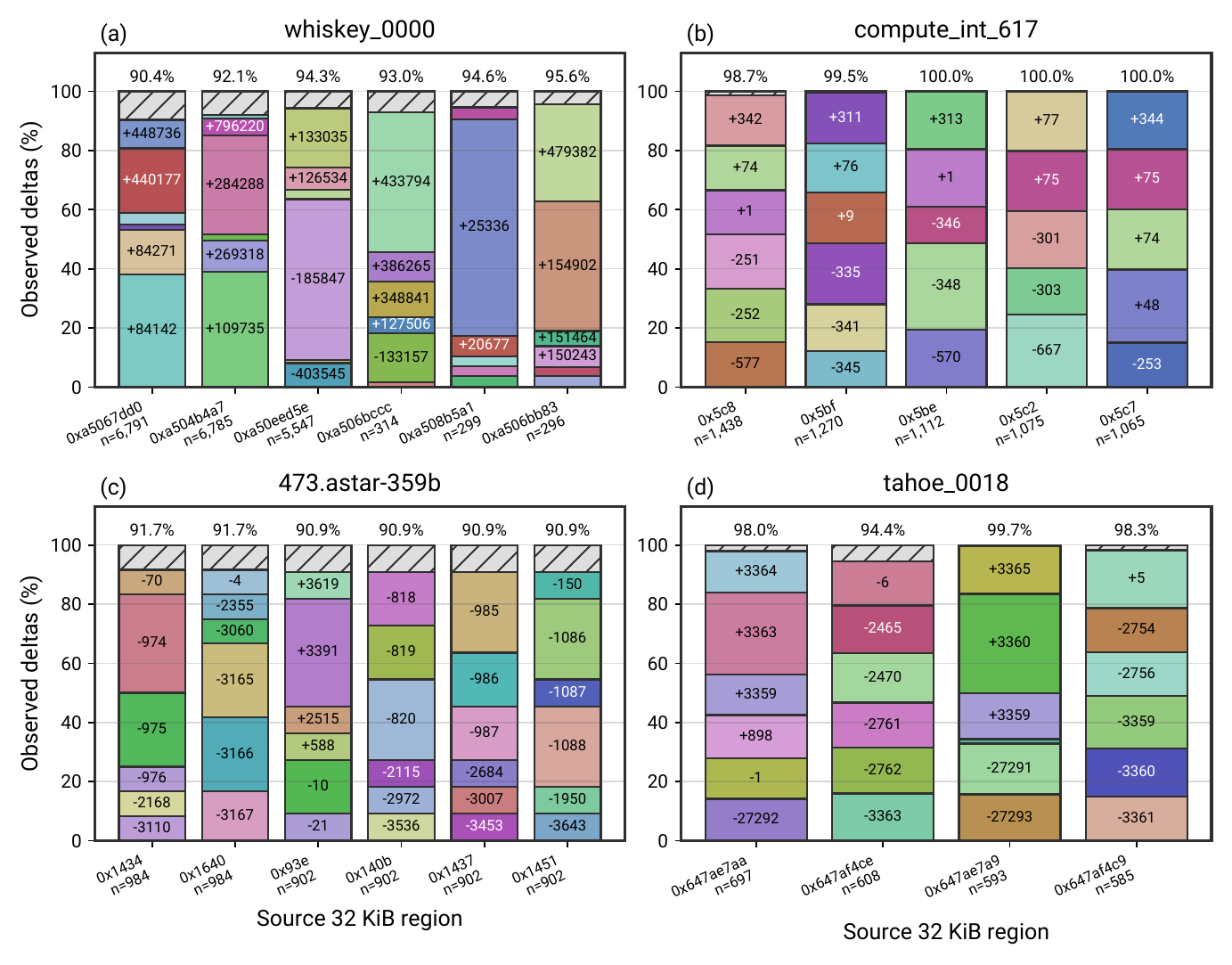}
  \caption{Per-PC delta distributions across selected 32\,KB source regions across four example workloads.}
  \label{fig:motivation_patterns}
\end{figure}

\textbf{Example per-PC delta distributions.}
\Cref{fig:motivation_patterns} shows the per-PC delta distributions of selected 32\,KB source regions from four example workloads. Each panel represents one workload, and each bar represents one source region.

We make two observations.
First, a small set of deltas accounts for 90.4\%--100\% of the transitions in every region shown.
Second, the dominant deltas differ across regions of the same workload. For example, in \texttt{tahoe\_0018}, region \texttt{0x647ae7a9} predominantly uses positive deltas near $+3360$, whereas region \texttt{0x647af4c9} predominantly uses negative deltas near $-3360$.
A prefetcher therefore needs to learn deltas separately for each source region.

\textbf{Why does a source region have only a few destination regions?}
Two properties of irregular applications explain the findings in Observation~1. 
First, in many irregular applications (e.g., graph analytics and pointer-chasing workloads), a small set of load instructions repeatedly traverses a few large data structures. Each instruction accesses related elements in sequence (i.e.,  a vertex followed by its neighbors or a node followed by the node it points to).
Second, programs tend to place related elements close together in virtual memory. For example, popular graph analytics frameworks~\cite{graphbig,ligra,gapbs} store each vertex's neighbors contiguously in an edge array.
As a result, after a load instruction accesses one source region, its subsequent accesses tend to reach only a few destination regions. The same few deltas recur whenever the instruction revisits that source region.
Per-PC tracking isolates the transitions generated by each load instruction, whereas global tracking interleaves transitions from instructions that may traverse different data structures. Per-PC tracking therefore produces a smaller successor set for each source region (\Cref{fig:motivation_pc_cond}). This regularity is the region-level analogue of the address-value regularity that AVD prediction~\cite{mutlu2005address} exploits in linked data structures.

\begin{figure}[h!]
  \centering
  \includegraphics[width=1.0\linewidth]{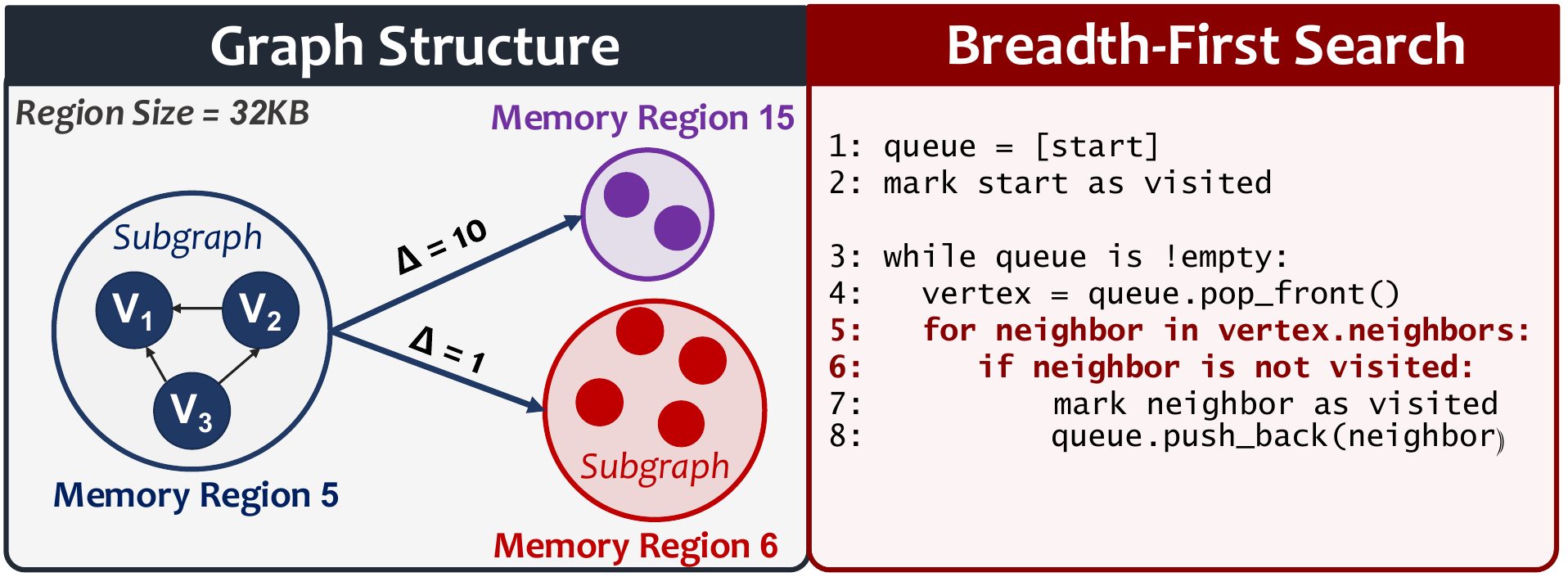}
  \caption{Example of how graph locality creates recurring deltas during BFS. (Left) Vertices organized in memory-adjacent subregions, yielding deltas $+1$ and $+10$. (Right) BFS pseudocode showing the load instructions that trigger PTWs.}
  \label{fig:graph_walk_example}
  \vspace{-1mm}
\end{figure}

\Cref{fig:graph_walk_example} uses breadth-first search (BFS) as an example graph traversal algorithm to illustrate how repeated accesses by the same load instruction and the placement of related data in nearby memory regions create a small set of recurring per-PC region deltas.
The left panel shows vertices $V_1$, $V_2$, and $V_3$ stored in region 5, with neighboring vertices packed into regions 6 and 15.
This layout arises because each vertex's neighbor list is stored contiguously, and vertices that are close in the graph often have nearby IDs, placing their data close together in virtual memory. The right panel shows the pseudocode for breadth-first search (BFS). For each vertex removed from the queue, BFS examines its neighbors (line 5) and enqueues those not yet visited (line 6). The load instructions on lines 5 and 6 trigger the page table walks.

For successive PTWs from the same PC that access region 5 and then region 6 or 15, the per-PC delta is $+1$ or $+10$, respectively.
The key is that the delta identifies a destination region, not an individual vertex.
Any transition from region 5 to a vertex in region 6 uses the same delta $+1$ while transitions to vertices in region 15 share delta $+10$.
Thus, if these transitions recur, retaining just two per-PC deltas for this source region would identify the likely destination regions without predicting the exact next vertex.
This example illustrates why a few per-PC deltas per source region capture most source-to-destination transitions even though the application accesses a potentially large number of memory regions.

\textbf{Observation~2: The size of the memory region associated with a delta creates a tradeoff between prediction coverage and prefetch accuracy.}
We define the \emph{region granularity} as the amount of contiguous virtual address space represented by one region. A 4\,KB region contains one virtual page, whereas a 32\,KB region contains eight consecutive 4\,KB pages. Changing the granularity changes how we group pages into source and destination regions and, therefore, the deltas between those regions.
Region granularity creates a potential tradeoff. A coarser granularity can increase prediction coverage because transitions to nearby pages collapse into the same destination region and share one delta.
However, a larger predicted region contains more pages. Prefetching its translations is therefore more likely to fetch PTEs for pages that the application never accesses, reducing prefetch accuracy and wasting memory bandwidth.
Conversely, a finer granularity limits overfetch but can split nearby destinations across several deltas and reduce the coverage of a fixed number of deltas.

\begin{figure}[h!]
  \centering
  \includegraphics[width=1.0\columnwidth]{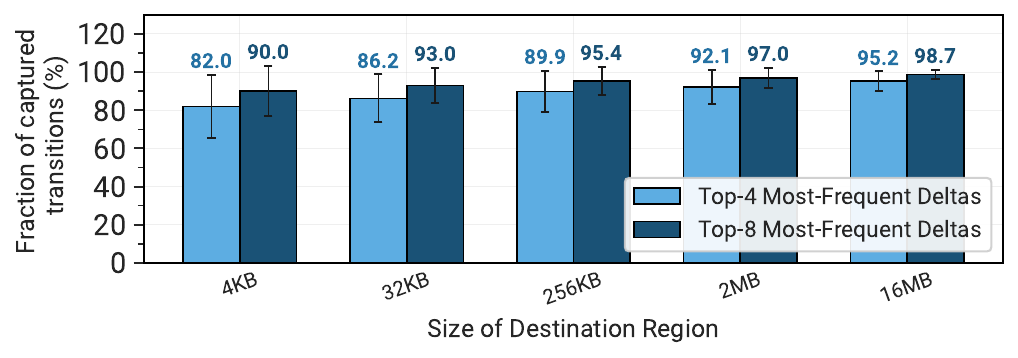}
  \caption{Average fraction of transitions captured by the four or eight most frequent per-PC deltas for each virtual memory region using different region sizes, across 200 workloads.}
  \label{fig:motivation_granularity}
\end{figure}

\Cref{fig:motivation_granularity} quantifies how region size affects the fraction of source-to-destination transitions captured.
For each evaluated region size, we partition the virtual address space into regions of that size and compute the deltas between successive page table walks triggered by the same PC. For each source region, we retain its four or eight most frequent deltas. We define \emph{coverage} as the fraction of source-to-destination transitions whose delta is in the set of retained (and most frequent) deltas.
We make two observations. First, increasing the granularity from 4\,KB to 32\,KB increases coverage from 82.0\% to 86.2\% with four deltas and from 90.0\% to 93.0\% with eight. Second, increasing the granularity further continues to increase coverage: 16\,MB regions reach 95.2\% with four deltas and 98.7\% with eight.

\begin{figure}[h!]
  \vspace{-1mm}
  \centering
  \includegraphics[width=1.0\columnwidth]{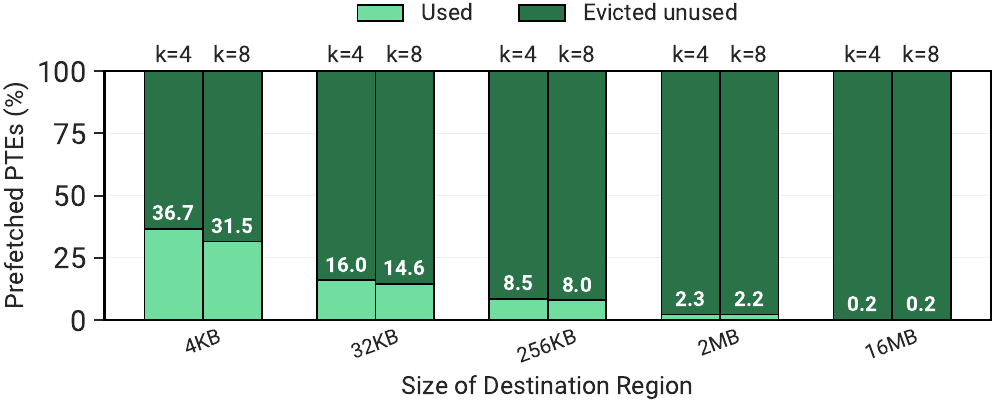}
  \caption{Fraction of prefetched PTEs that are used versus unused as the virtual memory region size increases, across 200 translation-intensive workloads.}
  \label{fig:motivation_granularity_accuracy}
\end{figure}

To quantify the cost of overfetching, \Cref{fig:motivation_granularity_accuracy} shows the used and unused prefetched PTEs (i.e., prefetch accuracy) in a simulated 1.5K-entry TLB, across 200 translation-intensive workloads. We classify a PTE as \emph{used} if a demand access uses its translation before the TLB evicts it, and as \emph{unused} otherwise.
We observe a rapidly compounding increase in unused prefetched PTEs as the region size increases. Each granularity step increases the number of PTEs prefetched for a destination region eightfold, while the fraction of prefetched PTEs that are unused also increases exponentially. This trend explains why the additional coverage of larger regions does not necessarily improve performance: larger regions increasingly consume TLB capacity with translations that are evicted before use.

\takeawaybox[2]{Region granularity creates a tradeoff between prefetch coverage and prefetch accuracy. Larger regions group nearby destination pages, allowing a fixed number of deltas per source region to capture more source-to-destination transitions. However, prefetching the translations of larger regions increases the fraction of unused prefetched PTEs, reducing prefetch accuracy.}

\begin{figure}[h!]
  \centering
  \includegraphics[width=1.0\columnwidth]{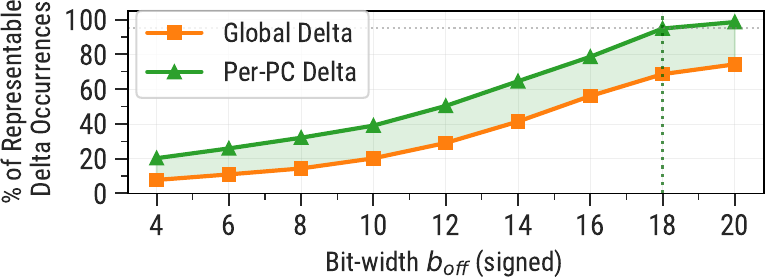}
  \caption{Average fraction of global and per-PC transition occurrences whose delta fits in a signed value of each bit width, using 32\,KB regions across 200 workloads.}
  \label{fig:motivation_bits}
\end{figure}

\textbf{Observation~3: Most per-PC region deltas fit in 18 bits.}\label{obs:encoding}
To determine how many bits each delta requires, \Cref{fig:motivation_bits} shows the average fraction of transition occurrences (i.e., $\langle\text{memory region}, \text{delta}\rangle$ pairs) whose signed delta fits in each bit width. We evaluate both global and per-PC tracking using 32\,KB regions across 200 workloads. 
We make two observations. First, per-PC deltas are more compact than global deltas: an 18-bit signed value can represent 94.9\% of per-PC delta occurrences, compared with 68.7\% of global delta occurrences. This is because per-PC tracking avoids the large deltas that can arise when the global stream interleaves unrelated instructions. Thus, per-PC tracking not only captures more transitions with a few deltas (Observation~1), but also reduces the number of bits needed to encode those deltas. Second, 16 bits represent 78.6\% of per-PC delta occurrences, while 18 bits can represent 94.9\%. At a 32\,KB region granularity, an 18-bit signed delta spans approximately $\pm$128K regions, or $\pm$4\,GB of virtual address space.

\takeawaybox[3]{Per-PC tracking enables compact delta storage: 18 bits represent 94.9\% of per-PC delta occurrences, so four delta fields require only 72 bits per source region.}

\section{TRAIL: Key Idea and Overview}
\label{sec:key_idea}

We introduce \chiron{}, whose key idea is to store deltas directly in unused bits of a virtual memory region's leaf PTE. This allows \chiron{} to 
(i) record deltas without requiring additional dedicated hardware-based tables or extra memory storage  and (ii) retrieve them efficiently during subsequent page table walks without extra memory accesses.

\begin{figure}[h!]
    \centering
  \includegraphics[width=1.0\columnwidth]{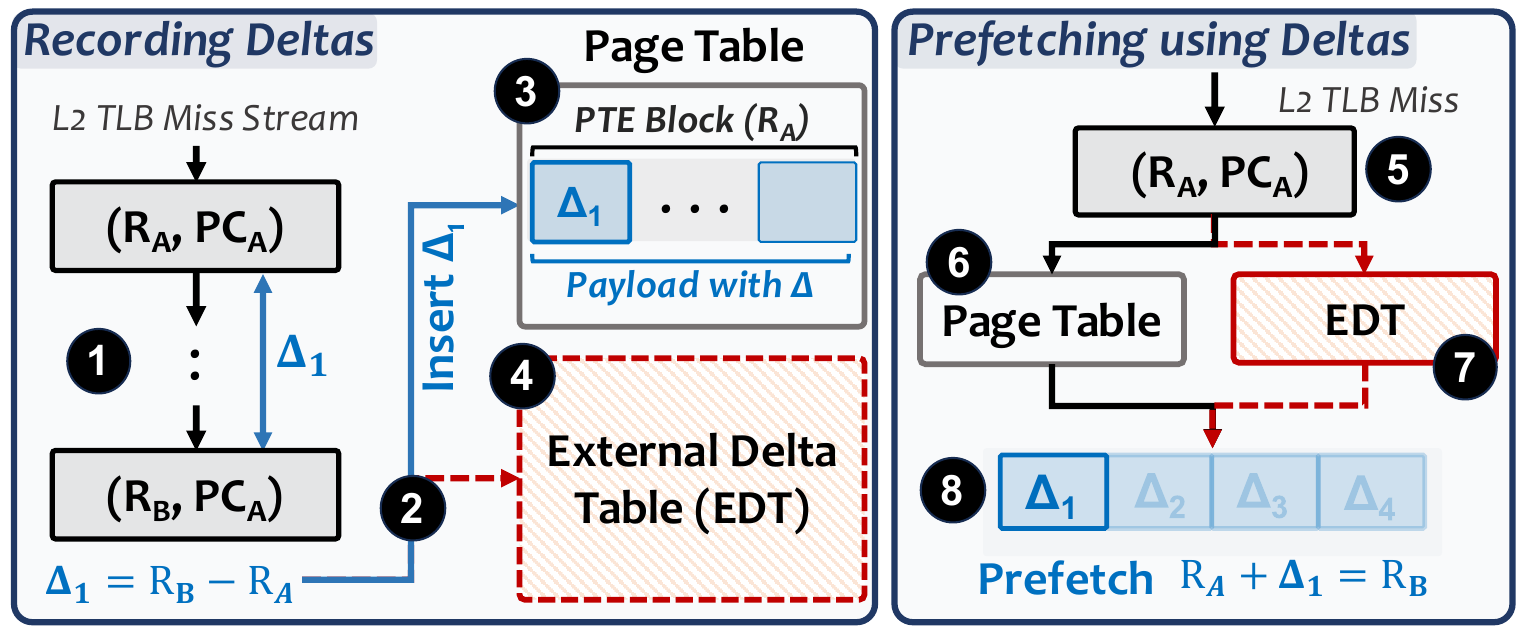}
  \caption{Overview of \chiron{}.}
    \label{fig:trail_overview}
\end{figure}

\Cref{fig:trail_overview} shows an overview of \chiron{}: (i) how it records deltas and stores them in the leaf PTEs of source regions and (ii) how it retrieves the stored deltas to prefetch the virtual-to-physical mapping of destination regions.

\textbf{Recording Deltas.}
\chiron{} uses a small hardware \emph{PC table} to remember the last region that triggered a demand PTW from each tracked instruction. In \Cref{fig:trail_overview}, successive misses from $\text{PC}_A$ access pages $\text{VPN}_A$ and $\text{VPN}_B$ in regions $R_A$ and $R_B$~(\blackcircled{1}). Using $R_A$ from the PC table, \chiron{} computes $\Delta_1=R_B-R_A$ ~(\blackcircled{2}) and records it at the PTE block corresponding to $R_A$~(\blackcircled{3}), since the delta describes where an access to that source region may lead next. Each region retains a small set of deltas and their confidence counters, which we call the region's \emph{payload}. These counters reflect how often deltas recur, allowing \chiron{} to retain frequently observed deltas and discard stale ones.
\chiron{} stores the payload in unused bits of the region's leaf PTEs under a best-effort hardware/software contract: the OS or hypervisor reserves only bits that it can safely repurpose. If the available bits cannot hold the desired number of deltas, \chiron{} can instead store the payload in an OS-managed external  delta table (EDT) (\blackcircled{4}). EDT is capable of retaining more deltas than the conventional PTE block permits, at the cost of additional storage and an additional lookup which occurs in parallel to the demand page table walk (\S\ref{sec:mech:encoding}).

\textbf{Prefetching.}
A later L2 TLB miss for $\text{VPN}_A$~(\blackcircled{5}) triggers a page table walk that fetches the leaf PTE cache block for region $R_A$~(\blackcircled{6}). If the PTE bits hold the payload, the page table walk retrieves it together with the PTEs and requires no additional memory access. If these bits are unavailable, \chiron{} instead retrieves the payload from the EDT~(\blackcircled{7}) by issuing a lookup in parallel with the page table walk. In the illustrated configuration, the eight PTEs map one 32\,KB region and each contributes 10 reserved bits to an 80-bit payload containing four deltas and their confidence counters. \chiron{} selects confident deltas, such as the recorded $\Delta_1$~(\blackcircled{8}), and adds them to the current region ID to prefetch the virtual-to-physical mapping of destination regions. In this example, $\text{R}_A+\Delta_1$ points to $\text{R}_B$, and \chiron{} prefetches its PTE.

\textbf{Key Benefits.}
Storing deltas in the source region's leaf PTEs provides three benefits.
First, the space for storing deltas scales with the application's memory footprint. The OS already allocates leaf PTEs to map the application's virtual memory. Thus, each source region has PTEs in which to store its deltas, and the space for deltas grows with the memory footprint, whereas a fixed-capacity hardware table cannot retain history for an increasing number of regions.
Second, \chiron{}'s hardware cost is low, since the deltas occupy bits that already exist but are unused, rather than requiring a dedicated hardware table or additional memory.
Third, \chiron{} retrieves the whole payload (i.e., multiple deltas) with low latency and no additional memory accesses, since the page table walk that fetches a region's PTEs also fetches its payload. Thus, \chiron{} can immediately use the retrieved deltas to determine which destination regions to prefetch.

\section{TRAIL: Detailed Design}
\label{sec:mechanism}

\chiron{} stores a payload of recorded deltas with each leaf page table block, while interpreting the deltas in that payload at a configurable prediction granularity. We describe five key aspects of its design:
(i)~the payload format and its storage in the page table or an external delta table (\S\ref{sec:mech:encoding}),
(ii)~the recording of deltas and management of their confidence (\S\ref{sec:mech:learning}),
(iii)~the use of recorded deltas to issue and chain prefetches (\S\ref{sec:mech:prediction}),
(iv)~the support for huge pages (\S\ref{sec:mech:implementation}), and
(v)~the PC table extensions for stride detection and preservation of evicted learning state (\S\ref{sec:mech:pc_extensions}).

\subsection{Delta Payload Organization}
\label{sec:mech:encoding}

\hspace*{\parindent}\textbf{Payload contents.} 
\chiron{} stores each payload in bits distributed across the PTEs of a single 64-byte leaf-PTE cache line. With 4\,KB pages, these eight PTEs map a contiguous 32\,KB virtual-address span, which we call a \emph{payload block}. A demand page table walk retrieves the payload in the same cache-line access that fetches the requested leaf PTE.
The payload-block size determines how prediction metadata is grouped, whereas the \emph{prediction-region} size determines the unit in which deltas are measured. These sizes can differ: a payload associated with a 32\,KB payload block can store deltas describing transitions between 128\,KB prediction regions. Each stored delta $\Delta$ is a signed difference between prediction-region IDs and predicts a transition from source region $R$ to destination region $R+\Delta$. 
Each payload contains $k$ pairs $(\Delta_i,c_i)$, where $\Delta_i$ is a signed prediction-region delta and $c_i$ is its saturating confidence counter. Our default payload contains four pairs, each consisting of an 18-bit two's-complement delta and a 2-bit confidence counter, for a total of $4\times(18+2)=80$ bits per payload block. Each delta can encode an offset in $[-2^{17},2^{17}-1]$, measured in prediction regions.

\textbf{Available capacity.} ISAs distinguish PTE bits that encode architectural state from bits that are ignored by hardware, available to software, or reserved. The assignments differ across architectures and can depend on the physical-address width and enabled features~\cite{intelx86manual,amdmanual,arm2013}.
Embedding deltas in the PTE is therefore a best-effort deployment choice and not a universal assumption. The OS or hypervisor supplies the available capacity to \chiron{}, which leaves every bit outside its assigned payload unchanged.
In \S\ref{sec:eval:pte_delta_budget}, we evaluate \chiron{} under different PTE bit budgets and compare it against \chiron{}-External, which stores the same logical payload in an external delta table (EDT) that resides in main memory.

\subsubsection{In-PTE Storage and Capacity}
When embedding deltas in the PTE is possible, each payload block's metadata is distributed across the eight leaf PTEs that share the corresponding 64-byte cache line. Our default x86-64 configuration assumes that the hardware/software contract provides 10 bits per leaf PTE, enough to hold one 80-bit payload across that block. This ensures that (i) no additional hardware storage or memory capacity is required to store the payload and (ii) a demand page table walk to any page in the block retrieves the payload along with the PTEs, without a separate metadata read (i.e., an extra memory access).

\subsubsection{External Delta Table (EDT)}
\label{sec:mech:external}

When embedding deltas in the PTE is not possible or insufficient, \chiron{}-External stores the same payload in an OS-managed external delta table (EDT) that resides in main memory. Both in-PTE and EDT organizations associate one payload with each 32\,KB payload block (with 4\,KB pages). They only differ in how \chiron{} locates that payload, not in the deltas it records or the confidence rules it applies.

\textbf{Organization.} EDT is a three-level radix tree with 9 index bits per level. The two upper levels hold pointers, and the leaf level holds one $B$-bit payload per payload block. A 64-byte leaf cache line packs $\lfloor512/B\rfloor$ consecutive entries, i.e., six payloads of the default width. Consecutive payload blocks share a leaf line, and neighboring blocks share upper-level nodes. 

{\textbf{Allocation.} The payload storage cost is 10 bytes per 32\,KB payload block, i.e., 0.03\% of the mapped memory. The OS allocates physical frames for the EDT through the same allocator path that it uses for page tables and exposes its base address to the hardware using a dedicated register, similar to CR3 for x86-64 page tables~\cite{intelx86manual}.}

{\textbf{Lookup.} On a demand L2 TLB miss, \chiron{} looks up the payload associated with the demand page's payload block in the EDT in parallel with the demand page table walk. The EDT traversal reuses the existing page-walk caches (PWCs) for its upper-level nodes and retrieves the leaf block through the data cache hierarchy. The demand page table walk proceeds independently and does not wait for the EDT lookup. \chiron{} issues prefetches once the payload arrives.}

\subsection{Recording Deltas}
\label{sec:mech:learning}

\begin{figure}[h!]
	\centering
	\includegraphics[width=1.0\linewidth]{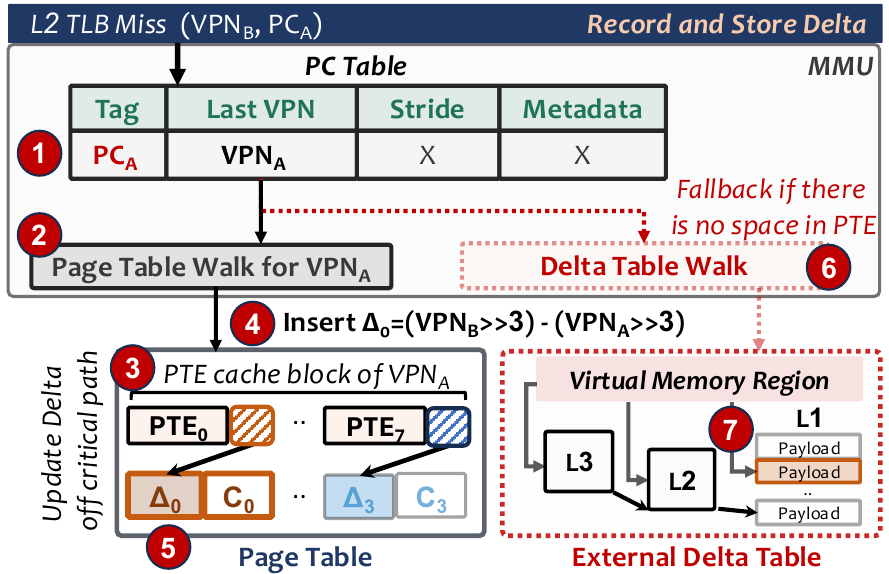}
	\caption{Workflow of recording and storing deltas in \chiron{}.}
	\label{fig:record_deltas}
\end{figure}

{To record a delta between two prediction regions, \chiron{} (i)~uses a small \emph{PC table}, as part of the processor's memory management unit (i.e., the hardware component responsible for address translation), which tracks, for each PC, the VPN that last triggered a PTW, and (ii)~updates the payload of the payload block that contains that preceding VPN with the computed delta. \Cref{fig:record_deltas} shows this process for an L2 TLB miss to $\text{VPN}_B$ triggered by $\text{PC}_A$, whose previous miss was to $\text{VPN}_A$.}

{\textbf{Find the preceding access.} After an L2 TLB miss, \chiron{} hashes $\text{PC}_A$ to index the PC table and compares the corresponding entry's tag with $\text{PC}_A$. 
In the example of \Cref{fig:record_deltas}, the entry matches and holds $\text{VPN}_A$ as the last VPN that triggered a PTW for $\text{PC}_A$~(\redcircled{1}).
\chiron{} then updates the entry's last VPN to $\text{VPN}_B$, so that the next miss from $\text{PC}_A$ learns a transition out of $\text{VPN}_B$.
On a tag mismatch, \chiron{} replaces the entry with the current PC (e.g., $\text{PC}_A$) and VPN (e.g., $\text{VPN}_B$) without recording a delta.}

{\textbf{Locate the source payload.} In \Cref{fig:record_deltas}, \chiron{} observes a transition from the source prediction region containing $\text{VPN}_A$ to the destination prediction region containing $\text{VPN}_B$. It records that transition in the payload block that contains $\text{VPN}_A$.
To fetch this source payload, \chiron{} walks the page table~(\redcircled{2}) and fetches the leaf PTE cache block that contains $\text{VPN}_A$~(\redcircled{3}).
Fetching the source payload (i) is performed in parallel and off the critical path of the current demand page table walk (for $\text{VPN}_B$) and (ii) typically benefits from cache locality due to the recent access of the same page table nodes and leaf PTE cache block by the earlier demand walk for $\text{VPN}_A$.}

{\textbf{Compute and insert the delta.} Since each prediction region contains $2^r$ pages, \chiron{} obtains the prediction region containing a VPN by removing the VPN's $r$ least significant bits (i.e., by shifting the VPN right by $r$ bits). It then computes the delta from the source prediction region to the destination prediction region as
$\Delta_0=(\text{VPN}_B \gg r)-(\text{VPN}_A \gg r)$
where $\text{VPN}_A \gg r$ is the source prediction region and $\text{VPN}_B \gg r$ is the destination prediction region (\redcircled{4}). \chiron{} gathers the payload bits from $\text{PTE}_0$ through $\text{PTE}_7$ of the source payload block (\redcircled{3}) and decodes the delta and confidence slots (\redcircled{5}). If a slot already contains $\Delta_0$, \chiron{} increments its confidence level. Otherwise, \chiron{} places $\Delta_0$ in a free slot or replaces a low-confidence delta, as described in \S\ref{sec:mech:confidence}. The update changes only the payload bits and marks the PTE cache block as dirty. The cache hierarchy writes the block back to memory when it is evicted.}

{\textbf{EDT lookup and delta insertion.} When the payload cannot be embedded in the PTEs, \chiron{}-External uses the source payload block number to perform a delta table walk through the three EDT levels (\redcircled{6}) and locate the source payload (\redcircled{7}). It then inserts $\Delta_0$ and updates its confidence using the same procedure as the in-PTE design.}

\subsubsection{Confidence Management}
\label{sec:mech:confidence}

Confidence determines which deltas remain in the payload and which deltas can trigger prefetches. When an observed delta $\Delta$ matches a payload slot (i.e., $\Delta$ is already recorded in the payload), \chiron{} increments its confidence counter up to $c_\text{max}=2^{b_\text{conf}}-1$. If no slot matches, \chiron{} replaces the slot with the lowest confidence and initializes the new pair to $(\Delta,c_\text{init})$. To prevent old deltas from remaining indefinitely in the payload, an aging policy reduces the confidence of other slots when a match occurs or before replacing a slot after a miss. A newly inserted delta can trigger a prefetch on the next visit to its source payload block if $c_\text{init}\geq c_\text{th}$. Otherwise, \chiron{} must observe it again until its confidence reaches the issue threshold $c_\text{th}$.

\subsection{Prefetching using Recorded Deltas}
\label{sec:mech:prediction}

\begin{figure}[h!]
	\centering
	\includegraphics[width=1.0\linewidth]{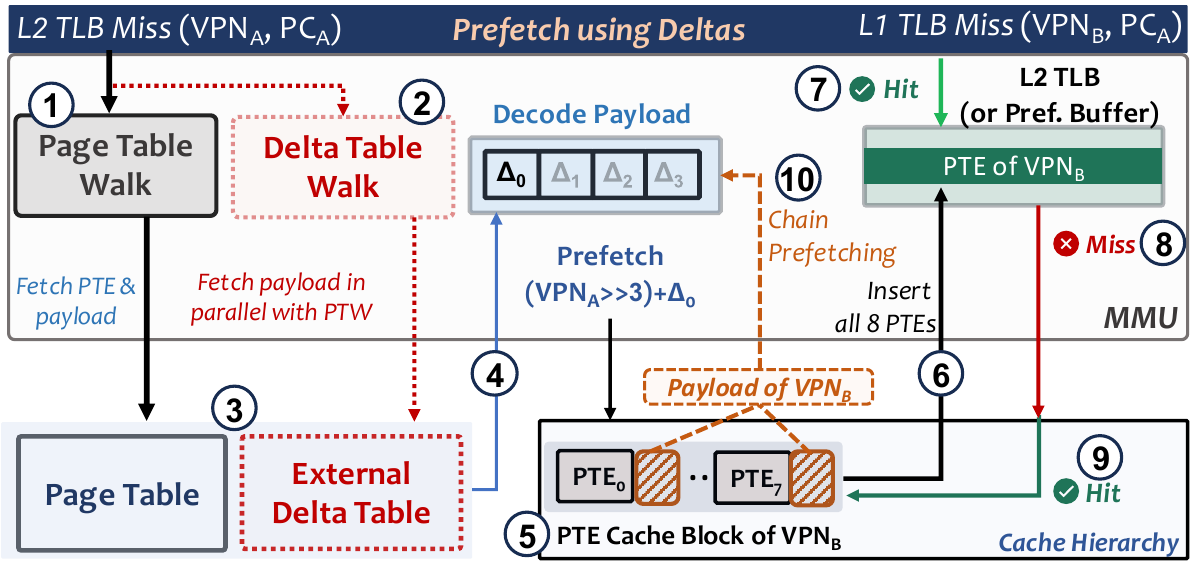}
	\caption{Workflow of prefetching using deltas in \chiron{}.}
	\label{fig:prediction}
\end{figure}

\Cref{fig:prediction} illustrates the workflow of prefetching using deltas in \chiron{}. We follow the same example as in \Cref{fig:record_deltas}.

\textbf{Fetch deltas.} On a demand L2 TLB miss to $\text{VPN}_A$, the page table walk fetches the leaf PTE block that contains $\text{VPN}_A$. For in-PTE storage, that block also provides the embedded payload of the current 32\,KB payload block~(\circled{1}). With \chiron{}-External, an EDT walk instead retrieves the same payload in parallel with the demand page table walk (\circled{2}). In both cases, the payload contains deltas interpreted relative to the current prediction region, $R_A=\text{VPN}_A\gg r$, and points to likely destination regions~(\circled{3}).

\textbf{Select deltas.} \chiron{} decodes the payload and examines its slots in descending confidence~(\circled{4}). It rejects zero deltas (i.e., self-transitions) and counters below the issue threshold $c_\text{th}$. For each remaining slot, it computes the target region $R'=R_A+\Delta_i$ and its base VPN: $R'\ll r$. In the example shown in \Cref{fig:prediction}, $\Delta_0$ is used to prefetch the prediction region that contains $\text{VPN}_B$. 

\textbf{Fetch translations.} Once the current payload arrives and \chiron{} selects a confident delta, it computes the target region $R'$ and immediately starts a page table pre-walk for that region. The target region covers VPNs $(R'\ll r)+p$ for $0\leq p<2^r$. The pre-walk fetches their translations and brings the required leaf PTE cache blocks into the cache hierarchy (\circled{5}).
The number of blocks fetched for a prediction depends on the configured prediction granularity because each 64-byte block holds eight 8-byte PTEs. With 4\,KB pages and $r=3$, the predicted destination region contains eight pages, so one leaf PTE block covers it. Larger prediction regions require more blocks. For example, the 128\,KB prediction region that \chiron{}-Best uses (\Cref{tab:evaluated_systems}) contains 32 pages and spans four leaf PTE blocks. The payload placement does not change: the source payload still resides in one 32\,KB leaf-PTE block and arrives with one leaf-block access, but the selected delta now identifies a coarser 128\,KB destination region. \chiron{} therefore fetches every leaf PTE block that covers the selected destination region. When these prefetch walks complete, \chiron{} places all fetched PTEs in the L2 TLB (or in a prefetch buffer) (\circled{6}).

\textbf{Prefetching benefits.}
 On a later L1 TLB miss to $\text{VPN}_B$, a hit in the L2 TLB (or prefetch buffer) directly supplies the virtual-to-physical mapping without performing a PTW (\circled{7}). If the prefetched PTE has been evicted before the demand memory access is performed and \chiron{} does not find it in the L2 TLB or prefetch buffer (\circled{8}), a conventional demand PTW is performed. Even then, the demand PTW may still find the required PTE in the cache hierarchy (which would otherwise be fetched from the main memory) (\circled{9}), reducing the PTW and address translation latency.
\subsubsection{Chained Prefetching}
\label{sec:mech:chaining}

\chiron{} uses fetched payloads in a recursive manner to prefetch the PTEs of subsequent prediction regions. For example, once the prefetch for $\text{VPN}_B$'s prediction region completes, \chiron{} examines the payload of every fetched payload block in that region and uses each of them to continue prefetching (\circled{10}). A FIFO queue holds the retrieved payloads for processing, so \chiron{} explores immediate successors before more distant ones. Thus, if a predicted region spans multiple payload blocks, \chiron{} enqueues all of their payloads and may extend the chain from any of them. A depth limit, $d_\text{max}$, bounds how far this process continues: a depth of zero allows only direct predictions from the demand region, while a depth of one allows one additional prediction step from those prefetched regions. \chiron{} tracks visited prediction regions, starting with the demand region, to avoid cycles and repeated prefetches within the same chain of predictions.
In an extended version of this paper~\cite{trail_extended} we explain in detail how the chaining mechanism evaluates and scores prediction paths to optimize prefetching efficiency.

\subsection{Support for Huge Pages}
\label{sec:mech:implementation}

\chiron{} supports 2\,MB huge pages by applying the same separation between payload blocks and prediction regions, as well as the same payload format and confidence rules, used for 4\,KB pages. The page size changes only the page table level that stores the payload and the amount of memory covered by each payload block and prediction region.

\textbf{Payload organization.} A page table walk for a 2\,MB page terminates at a page middle directory (PMD) entry, which stores the page's virtual-to-physical mapping~\cite{intelx86manual}. Eight consecutive PMD entries occupy one 64-byte cache block and cover 16\,MB; we treat this span as a huge-page payload block. Prediction deltas are still interpreted over regions of $2^r$ consecutive 2\,MB pages. Thus, when $r=3$ the payload block and the prediction region coincide at 16\,MB, whereas larger $r$ values make the prediction granularity coarser. When the OS or hypervisor assigns sufficient PMD bits to \chiron{}, it distributes one 80-bit payload across each PMD block. Otherwise, \chiron{} stores the payload in the external delta table.

\textbf{Recording deltas.} Each PC-table tag includes the page size. Therefore, an entry matches only when both the PC and page size match, ensuring that the preceding and current accesses use the same page size. This prevents \chiron{} from recording a delta between a 4\,KB page and a 2\,MB page. Thus, a delta is measured in prediction regions of the matching page size. For example, when $r=3$, a delta of one advances by 16\,MB for 2\,MB pages, compared to 32\,KB for 4\,KB pages.

\textbf{Prefetching translations.} When a demand page table walk for a 2\,MB page retrieves the PMD block, \chiron{} decodes the payload of the current huge-page payload block. For each delta whose confidence reaches the issue threshold, \chiron{} computes the destination huge-page prediction region and pre-walks the page table to fetch all PMD entries that cover that region. Thus, huge-page prefetching follows the same steps as 4\,KB-page prefetching, but operates on PMD entries and on huge-page prediction regions whose size is determined by $r$.

\subsection{Extensions to \chiron{}}
\label{sec:mech:pc_extensions}

{We extend \chiron{} in two ways: (i)~ we enhance the PC-table to predict regular address translation patterns and improve prefetch accuracy and coverage (\S\ref{sec:mech:stride}), and (ii)~propose virtualizing the PC table to preserve learning state when applications with large instruction working sets experience frequent PC-table evictions (\S\ref{sec:mech:virt_pc}).}

\subsubsection{Per-PC Stride Detection}
\label{sec:mech:stride}
\chiron{} augments the PC table with a conventional per-PC stride predictor to cover accesses for which the payload holds no useful delta history. 

\textbf{Training.} Each PC-table entry stores the last VPN, a candidate page stride $\delta=\text{VPN}_\text{curr}-\text{VPN}_\text{prev}$, and a confidence counter. Unlike region deltas $\Delta$, which relate prediction regions and reside in payload blocks, $\delta$ is measured in pages and remains in the PC table.  On every L2 TLB miss from a tracked PC, \chiron{} computes the new page stride, compares it against the candidate, increments confidence on a match up to the stride threshold, and otherwise replaces the candidate and resets confidence to one. It then saves the current VPN for the next comparison.

\textbf{Prefetching.} Once confidence reaches the threshold, \chiron{} prefetches translations for $\text{VPN}_\text{curr}+j\delta$, where $j=1,\ldots,D$ and $D$ is the prefetch degree. These prefetches target individual pages, not entire prediction regions, and require no recorded payload. While the stride remains stable, \chiron{} stops recording that PC's prediction-region deltas in the PTEs or EDT, but it continues using any existing payloads for delta-based prefetching.

\subsubsection{Virtualized PC Table}
\label{sec:mech:virt_pc}

Frequent PC-table evictions discard the previous VPN and stride state needed for recording deltas. To preserve this state for applications with large instruction footprints (i.e., large number of PCs issuing memory accesses that lead to PTWs) without enlarging the on-chip PC table, \chiron{} stores evicted entries in a memory-resident backing store and restores them when the corresponding PC appears again in the instruction stream. The on-chip PC table thus acts as a cache of per-PC learning state. An extended version of this paper~\cite{trail_extended} provides the quantitative motivation and the full description of the virtualized PC table.

\section{Evaluation Methodology}
\label{sec:methodology}

We evaluate \chiron{} using \virtuoso{}~\cite{kanellopoulos2025virtuoso,virtuoso_github}, a simulation framework that enables fast and accurate full system simulation and is validated against real systems, on top of \sniper{}, an event-based multicore simulator~\cite{sniper}. We extend our simulation infrastructure to model \chiron{} and all evaluated systems. \chiron{} is fully open source at \href{https://github.com/CMU-SAFARI/Virtuoso/tree/trail-artifact-release}{github.com/CMU-SAFARI/Virtuoso/TRAIL}.

\begin{table}[ht!]
  \centering
  \caption{Baseline simulated system configuration.}
  \label{tab:system_config}
  \scriptsize
  \renewcommand{\arraystretch}{1.05}
  \setlength{\tabcolsep}{3pt}
  \begin{tabularx}{\columnwidth}{|>{\raggedright\arraybackslash}p{0.28\columnwidth}||>{\raggedright\arraybackslash}X|}
    \hline
    \textbf{Component} & \textbf{Simulation Configuration} \\
    \hline
    \hline
    Core & 8-wide OoO; 512-entry ROB; x86-64; 5.1\,GHz  \\
    L1 I-cache & 64\,KB; 16-way; 4 cycles; pLRU \\
    L1 D-cache & 48\,KB; 12-way; 4 cycles; pLRU; IP-Stride~\cite{chen1995tc} \\
    L2 cache & 2\,MB; 16-way; 12 cycles; pLRU; Streamer~\cite{jouppi_prefetch} \\
    L3 cache (LLC) & 2\,MB/core or 8\,MB/core; 16-way; 36 cycles; pLRU \\
    DRAM & 128\,GB DDR4-2400; 4 channels; 4 ranks; 32 banks \\
    L1 I-TLB & 64 entries; 4-way; 1 cycle \\
    L1 D-TLB & 1x64-entry (4\,KB); 1x64-entry (2\,MB); 4-way, 1 cycle \\
    L2 TLB & 2048 entries; 8-way; 12 cycles; CHiRP~\cite{chirpMICRO2020} \\
    Page table walkers & 4 walkers; FCFS scheduling for demand/prefetch walks \\
    Page structure caches & 3 levels; 32 entries per level; 4-way; 2 cycles \\
    TLB prefetch buffer & 512 entries; 2-way; 6 cycles \\
    \hline
  \end{tabularx}
\end{table}

\textbf{Workloads.}
\Cref{tab:workloads} lists all the workloads used in our evaluation.
We collect 200 traces across multiple benchmark suites that stress the address translation subsystem (address translation consumes more than 3\% of total execution time).  We include datacenter workloads (Google~\cite{googletraces}), 
mobile SoC workloads (CVP-1~\cite{cvp1}), graph analytics (Graph~\cite{graphbig,gapbs,ligra}), CPU benchmarks (SPEC CPU~\cite{spec2017,spec2006}), and other memory-intensive applications (Huawei DPC-4 Paoding-kbest~\cite{dpc4}, XSBench~\cite{xsbench}, PARSEC Canneal~\cite{bienia2008parsec}). In single-core system evaluation, we execute each benchmark for 300M instructions. In four-core evaluation, we create 100 mixes of the top-50 most translation-intensive benchmarks (based on how much time they spend in address translation) and execute each mix for 50M instructions.

\begin{table}[t!]
  \centering
  \caption{Evaluated workloads.}
  \label{tab:workloads}
  \footnotesize
  \renewcommand{\arraystretch}{1.12}
  \setlength{\tabcolsep}{4pt}
  \begin{tabularx}{\columnwidth}{@{}>{\raggedright\arraybackslash}p{0.34\columnwidth}>{\raggedright\arraybackslash}Xr@{}}
    \toprule
    \textbf{Suite} & \textbf{Description} & \textbf{\#Traces} \\
    \midrule
    Google Server~\cite{googletraces} & Datacenter traces (9 service families) & 137 \\
    CVP-1~\cite{cvp1} & Qualcomm mobile SoC traces &  34 \\
    Graph Analytics~\cite{graphbig,gapbs,ligra} & BFS, BC, SSSP, PR, CC, TC & 12 \\
    SPEC CPU~\cite{spec2017,spec2006}       & SPEC CPU 2006/2017 (mcf, lbm, etc.)    &   9 \\
    PARSEC Canneal~\cite{bienia2008parsec} & Multi-threaded simulated annealing &  4 \\
    Huawei DPC-4~\cite{dpc4}  & Paoding-kbest traces                    &   3 \\
    XSBench~\cite{xsbench}    & Monte Carlo neutron transport            &   1 \\
    \midrule
    \textbf{Total}             &                                         & \textbf{200} \\
    \bottomrule
  \end{tabularx}
\end{table}

\textbf{Evaluated Systems.}
We compare \chiron{} against eight systems:
(i)~\textbf{No-TLB-Prefetcher}: the baseline without TLB prefetching,
(ii)~\textbf{IP-Stride~\cite{chen1995tc}}: the baseline with PC-based stride TLB prefetching,
(iii)~\textbf{Next-Page~\cite{Bala1994SoftwarePA}}: prefetches translations at fixed page offsets -1, -2, +1, and +2,
(iv)~\textbf{DP~\cite{kandiraju2002going}}: prefetches translations using recurring deltas stored in an on-chip table,
(v)~\textbf{Recency~\cite{saulsbury2000recency}}: prefetches translations using a memory-resident list ordered by L2 TLB eviction recency,
(vi)~\textbf{Berti~\cite{navarro2022berti}}: adapts Berti's data prefetching mechanism to prefetch translations,
(vii)~\textbf{ATP~\cite{vavouliotis2021}}: combines H2P (two recent global miss deltas), MASP (per-PC stride), and STP (fixed offsets $\pm1$ and $\pm2$), and uses a counter-based throttler to select among them or disable prefetching when recent predictions are not useful,
(viii)~\textbf{Perfect L2 TLB}: an L2 TLB access always hits and the hit latency is charged normally. \Cref{tab:evaluated_systems} lists each system's configuration in detail. All evaluated systems use the same processor, cache hierarchy, and data prefetchers. We enable the same IP-Stride TLB prefetcher in all evaluated TLB prefetching configurations except \chiron{}, which uses its own per-PC stride detector. All TLB prefetchers use the same prefetch buffer, and demand and prefetch walks share the same page table walkers with first-come, first-served scheduling (\Cref{tab:system_config}).

\textbf{\chiron{} Configurations.}
\Cref{tab:evaluated_systems} lists the settings of our Trail configurations: (i) \textbf{\chiron{}-Best} and (ii) \textbf{\chiron{}-External}.
Unless otherwise specified, we report \textbf{\chiron{}-Best}, the best-performing \chiron{} configuration. An extended version of this paper~\cite{trail_extended} presents the design space exploration from which we derive this configuration. \textbf{\chiron{}-External} uses the same settings as \textbf{\chiron{}-Best}, except that it stores payloads in the external delta table instead of embedding them in PTEs and uses 4-bit confidence counters (4 deltas $\times$ (18 delta bits + 4 confidence bits) = 88 bits per payload), which the EDT can accommodate.

\begin{table}[ht!]
  \centering
  \caption{Configuration of evaluated systems and \chiron{}.}
  \label{tab:evaluated_systems}
  \scriptsize
  \renewcommand{\arraystretch}{1.05}
  \setlength{\tabcolsep}{3pt}
  \begin{tabularx}{\columnwidth}{|>{\raggedright\arraybackslash}p{0.28\columnwidth}||>{\raggedright\arraybackslash}X|}
    \hline
    \textbf{Evaluated System} & \textbf{Configuration} \\
    \hline

    No-TLB-Prefetcher & No translation prefetching; Data prefetchers are enabled \\
    IP-Stride~\cite{chen1995tc} & PC-based stride TLB prefetching; Lookahead: 4; Degree: 3\\
    Next-Page~\cite{Bala1994SoftwarePA} & Prefetches VPNs at fixed page offsets -1, -2, +1, and +2 \\
    DP~\cite{kandiraju2002going} & 256-row, 4-way SRAM table; 8 delta slots per row \\ 
    Recency~\cite{saulsbury2000recency} & Memory-resident recency list with previous/next links; Prefetch for +1, -1, and -2 offsets in the recency list \\ 
    Berti~\cite{navarro2022berti} & Adapted for TLB prefetching; 1024-entry PC table; 256-entry record-page table; 10 offsets; Degree 3 \\ 
    ATP~\cite{vavouliotis2021} & H2P: two recent global miss deltas; MASP: per-PC stride; STP: page offsets $\pm1$/$\pm2$; counter-based selector/throttler can disable prefetching; 1024-entry MASP; 64-entry sampler; 16-entry FPQ \\ 
    Perfect L2 TLB & Always hits in the L2 TLB; 12-cycle hit latency \\ 
    \hline
    \hline
    \multicolumn{2}{|l|}{\textbf{\chiron{}-Best and \chiron{}-External Configurations}} \\
    \hline
    Predictor & 64-entry PC table \& virtualized; per-PC stride detection;\\
    Region and payload & Prediction granularity: 128\,KB regions ($r{=}5$); Payload storage: 4 deltas $\times$ (18 delta bits + 2 confidence bits) = 80 bits per 32\,KB leaf PTE block; \\
    Confidence/Chaining & Initial confidence: 3; Chaining depth: 3; \\
    \chiron{}-Best & Payloads embedded in PTEs (i.e., in-PTE \chiron{})\\
    \chiron{}-External & Same predictor settings as \chiron{}-Best; Payloads reside in an external delta table (\S\ref{sec:mech:external}); 4-bit confidence counters: 4 deltas $\times$ (18 + 4) = 88 bits per payload \\
    \hline
  \end{tabularx}
\end{table}

\begin{figure*}[b]
  \centering
  \includegraphics[width=0.95\textwidth]{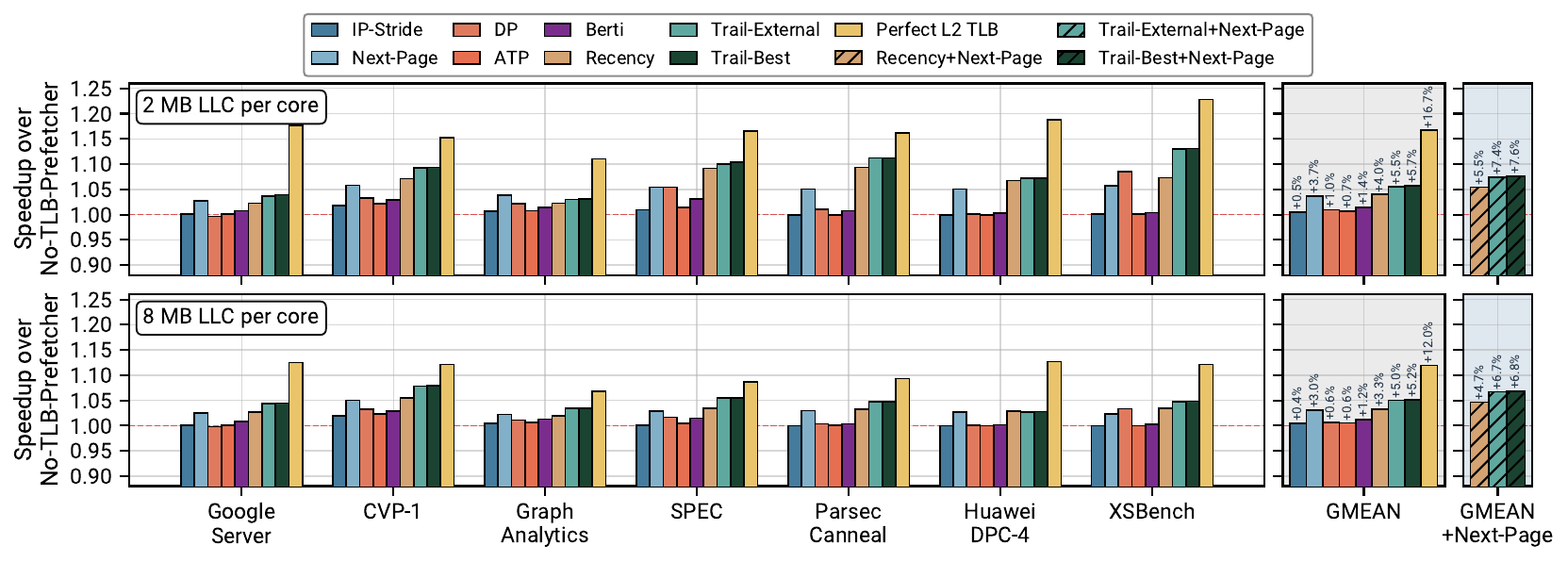}
  \caption{Average speedup provided by all evaluated prefetchers over No-TLB-Prefetcher for each workload family, with a 2\,MB LLC slice per core (top) and an 8\,MB LLC slice per core (bottom). The two rightmost panels show the geometric mean across all 200 workloads: (middle) shows the geometric mean for standalone prefetchers, and (right) shows the geometric mean for the configurations that combine Recency and \chiron{} with Next-Page.}
  \label{fig:prefetcher_comparison}
\end{figure*}

\section{Evaluation Results}
\label{sec:results}

We evaluate \chiron{} via four sets of experiments.
First, we evaluate \chiron{}'s performance in a single-core system, explain where \chiron{}'s performance benefit comes from, and quantify the overhead of updating \chiron{}'s deltas (\S\ref{sec:eval:speedup}-\S\ref{sec:eval:coverage}).
Second, we evaluate \chiron{}'s performance in a four-core system (\S\ref{sec:eval:multicore}).
Third, we study \chiron{}'s sensitivity to three system settings: large pages, DRAM bandwidth, and number \& width of deltas used by \chiron{} (\S\ref{sec:eval:system_sensitivity}).
Fourth, we quantify the contribution of each of \chiron{}'s components (\S\ref{sec:eval:design}).

\subsection{Single-core Performance}
\label{sec:eval:speedup}

\noindent\textbf{Performance Comparison.}
Figure~\ref{fig:prefetcher_comparison} shows (leftmost panel) the average performance improvement provided by each TLB prefetcher across each of the seven workload families, (middle panel) the geomean speedup provided by each TLB prefetcher across all 200 workloads and (rightmost panel) the geomean speedup provided by systems that combine Recency and \chiron{} with Next-Page across all 200 workloads, all normalized to No-TLB-Prefetcher in a system featuring a 2\,MB (top) and an 8\,MB (bottom) LLC slice per core. We make five key observations.

First, \chiron{}-Best provides the highest average performance improvement compared to all evaluated TLB prefetchers.
With a 2\,MB (8\,MB) LLC slice per core, \chiron{}-Best improves performance by 5.7\% (5.2\%) on average over No-TLB-Prefetcher.
The best-performing prior TLB prefetcher, Recency, improves performance by 4.0\% (3.3\%) on average, 1.7\% (1.8\%) less than \chiron{}-Best.

Second, \chiron{}-Best, i.e., the best-performing \chiron{} configuration,  outperforms prior prefetchers on most workload families with large, irregular translation streams.
This includes Google Server production-grade applications~\cite{googletraces}, our largest workload family (137 of the 200 workloads), on which \chiron{}-Best improves performance by 4.0\% (4.5\%) over No-TLB-Prefetcher and outperforms the best prior prefetcher, Recency, by 1.7\% (1.7\%).
These workloads repeatedly traverse pointer-based and large-footprint data structures.
Their address translation streams contain recurring but non-strided deltas (we demonstrate such patterns in \S\ref{sec:bg:motivation}), which \chiron{}'s per memory-region deltas capture effectively, but stride- and stream-based prefetchers do not.

Third, a larger LLC reduces the performance headroom of TLB prefetching, but \chiron{} still captures a larger fraction of the remaining headroom as the LLC grows.
When the LLC slice grows from 2\,MB to 8\,MB, perfect L2~TLB's improvement drops from 16.7\% to 12.0\%, since more page table walk requests hit in the LLC.
In contrast, \chiron{}-Best's improvement drops only from 5.7\% to 5.2\%, so \chiron{}'s share of the perfect L2~TLB's improvement increases from 34\% to 43\%.
This larger share has two causes.
First, \chiron{} effectively targets the page table walks that lead to DRAM accesses, which a larger LLC does not eliminate.
Second, a larger LLC keeps more of \chiron{}'s prefetched PTE cache blocks until the demand access uses them.
This effect is visible even in Google Server, one of the workload families where \chiron{}-Best's performance improvement increases with a larger LLC, from 4.0\% to 4.5\%.

Fourth, \chiron{}-External nearly matches \chiron{}-Best, which shows that most of \chiron{}'s benefit comes from its prediction scheme rather than from storing the payload as part of the PTEs. For a 2\,MB (8\,MB) LLC slice per core, \chiron{}-External improves performance by 5.5\% (5.0\%) over No-TLB-Prefetcher, only 0.1\% less than \chiron{}-Best's 5.7\% (5.2\%). Thus, storing \chiron{}'s metadata outside the PTEs preserves nearly all of the \chiron{}'s benefit.

Fifth, as we show in the rightmost panel of Figure~\ref{fig:prefetcher_comparison}, Recency+Next-Page improves average performance by 5.5\% (4.7\%), whereas \chiron{}-External+Next-Page and \chiron{}-Best+Next-Page improve performance by 7.4\% and 7.6\% (6.7\% and 6.8\%), respectively. Thus, combining Next-Page with \chiron{} yields the highest performance among all evaluated TLB prefetching techniques.

We draw three main conclusions. First, \chiron{}-Best is the best-performing standalone TLB prefetcher across both evaluated LLC sizes: it provides the highest average performance, its benefit is largest on workloads with large, irregular translation streams, and it captures a larger share of the remaining TLB prefetching headroom as the LLC grows. Second, \chiron{}-External preserves nearly all of this benefit, which shows that \chiron{}'s advantage comes primarily from its prediction scheme rather than from storing metadata in the PTEs. Third, combining \chiron{} with Next-Page provides the highest overall performance among all evaluated TLB prefetching techniques.

\begin{figure}[h!]
  \centering
  \includegraphics[width=\columnwidth]{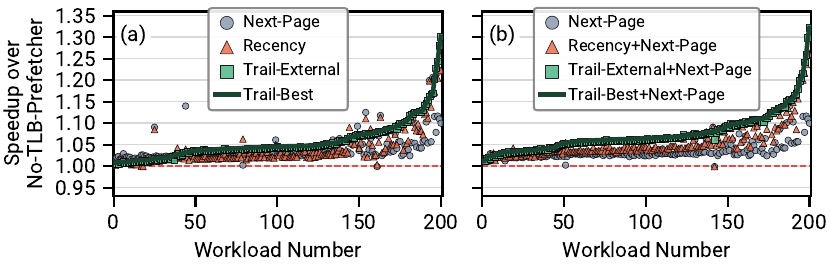}
  \caption{Per-workload speedup of the best performing prefetchers over No-TLB-Prefetcher across 200 workloads with a 2\,MB LLC slice per core: (a)~standalone Next-Page, Recency, \chiron{}-External, and \chiron{}-Best and (b)~all prefetchers combined with Next-Page.}
  \label{fig:scurve}
\end{figure}

\noindent\textbf{Per-workload Performance.}
To demonstrate how \chiron{}'s benefit varies across all 200 translation-intensive workloads, Figure~\ref{fig:scurve} shows per-workload speedup provided by (a) standalone Next-Page, Recency, \chiron{}-External, and \chiron{}-Best (the four best performing prefetchers) and (b) all prefetchers combined with Next-Page, over the No-TLB-Prefetcher baseline, in a system with a 2\,MB LLC slice per core.
We make three observations.
First, \chiron{}-Best improves performance by up to 30\% and 7.75\% over No-TLB-Prefetcher and Recency, respectively, and outperforms both across 151 of the 200 workloads.
Second, \chiron{}-Best does \emph{not} degrade performance on any of the 200 workloads.
Third, when combined with Next-Page,
 \chiron{}-Best outperforms both Next-Page and Recency+Next-Page on 173 of the 200 workloads. 
We conclude that: (i) \chiron{} provides large speedups on a large subset of workloads, (ii) \chiron{} never degrades performance on any evaluated workload, (iii) storing \chiron{}'s deltas in a separate data structure (\chiron{}-External) preserves nearly all of \chiron{}'s benefit, and 
(iv) combining Next-Page with \chiron{} yields the highest performance among all evaluated TLB prefetchers across almost all 200 workloads.

\subsection{Performance Analysis}
\label{sec:eval:coverage}

\begin{figure}[h!]
  \centering
  \includegraphics[width=\columnwidth]{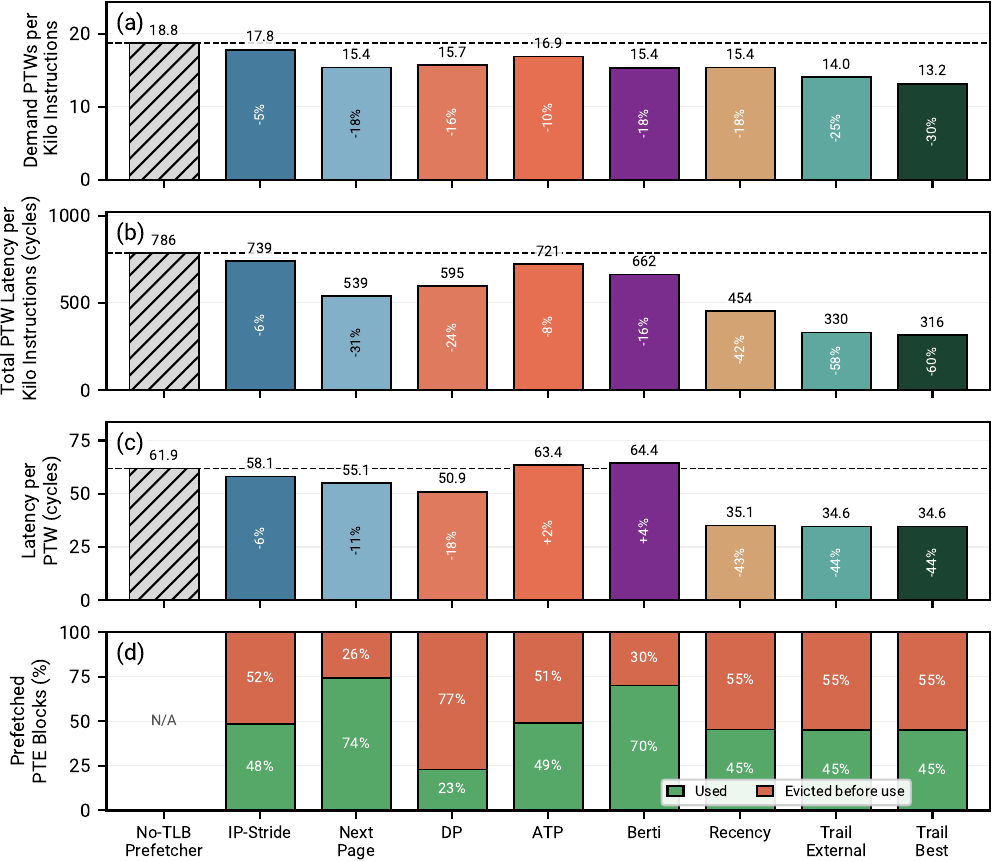}
  \caption{(a) Demand PTWs per kilo instructions, (b) total demand PTW latency in cycles per kilo instructions, (c) average latency per demand PTW in cycles, and (d) fraction of prefetched PTE cache blocks used or evicted before use, across all TLB prefetchers, for the 50 most translation-intensive workloads in a system with a 2\,MB LLC slice per core. Annotations in (a)--(c) show reductions relative to No-TLB-Prefetcher.}
  \label{fig:coverage_decomposition}
\end{figure}

\textbf{Coverage, Accuracy, and Translation Latency.}
To better understand \chiron{}'s effectiveness, we analyze how each prefetcher affects demand PTWs.
To highlight the properties of each prefetcher more distinctly, we perform this analysis on the 50 most translation-intensive workloads, for a 2\,MB LLC slice per core.
Figure~\ref{fig:coverage_decomposition} shows four metrics for each prefetcher:
(a)~the number of demand PTWs per kilo instructions,
(b)~the total demand PTW latency (in cycles) per kilo instructions,
(c)~the average latency per demand PTW (in cycles),
and (d)~the fraction of prefetched PTE cache blocks that are \emph{used} at least once or \emph{evicted} (from the LLC) before use.
We make three observations.

First, \chiron{}-Best and \chiron{}-External reduce total demand PTW latency per kilo instructions from 786 cycles to 316 (60\%) and 330 (58\%) cycles, respectively. The best-performing prior prefetcher, Recency, reduces this latency to 454 cycles (42\%), leaving 44\% more latency than \chiron{}-Best (454 versus 316 cycles).
Second, \chiron{}-Best's latency reduction comes from two sources.
(i)~\chiron{}-Best reduces the number of demand PTWs from 18.8 to 13.2 per kilo instructions, eliminating 30\% of demand walks compared to No-TLB-Prefetcher.
(ii)~\chiron{}-Best reduces the average latency per demand PTW from 61.9 to 34.6 cycles, 44\% lower compared to No-TLB-Prefetcher.
Recency nearly matches \chiron{}-Best's latency per remaining walk (35.1 versus 34.6 cycles), but does not eliminate as many demand walks as \chiron{}-Best (15.4 versus 13.2 per kilo instructions). Thus, \chiron{}-Best's advantage over Recency comes primarily from eliminating more demand walks.
Third, 45\% of \chiron{}-Best's and \chiron{}-External's prefetched PTE cache blocks are used at least once, matching Recency's accuracy. \chiron{}'s accuracy is lower than other prefetchers (e.g., Berti and ATP) since those prefetchers rarely detect stream/stride patterns and thus issue fewer prefetches overall.

We conclude that \chiron{} provides the largest reduction in address translation latency because it both eliminates the most page table walks and shortens the walks it does not avoid, without sacrificing accuracy compared to the best performing prefetcher (Recency).

\begin{figure}[h!]
  \centering
  \includegraphics[width=\columnwidth]{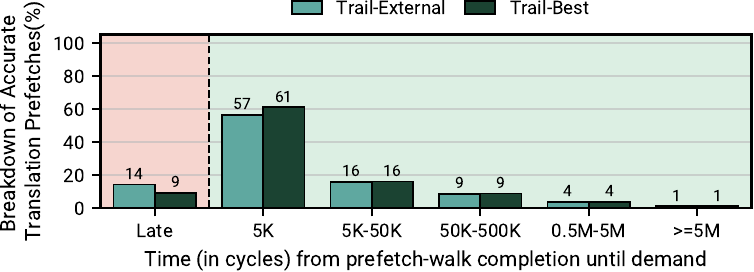}
  \caption{Distribution of average lead time (from prefetch-walk completion to demand access) of \chiron{}'s accurate prefetches, for a 2\,MB LLC slice per core, across all 200 workloads.}
  \label{fig:timeliness}
\end{figure}

\vfill\null
\noindent\textbf{Timeliness.}
Timeliness determines how much latency a prefetcher can hide.
A late prefetch, i.e., one that completes \emph{after} the demand access, hides only part of the walk latency, whereas a prefetch that arrives \emph{too early} risks eviction from the TLB (and caches) before the demand access uses it.
To quantify timeliness, we measure the \emph{lead time} of each accurate prefetch (i.e., a prefetched translation that was used by a subsequent demand access): the number of cycles from completing the prefetch walk (PTW) to the demand access to the prefetched translation.
Figure~\ref{fig:timeliness} shows the lead-time distributions of \chiron{}-Best's and \chiron{}-External's used prefetches.
The leftmost group (Late) contains prefetches that complete after the demand access, and the remaining groups bucket prefetches by how early they arrive.
We make two observations.
First, most of \chiron{}'s accurate prefetches complete before they are needed: 91\% of \chiron{}-Best's accurate prefetches and 86\% of \chiron{}-External's complete before the demand access.
Second, most of these timely prefetches do not arrive far in advance. 61\% of \chiron{}-Best's accurate prefetches and 57\% of \chiron{}-External's complete fewer than 5K cycles before the demand access, so the prefetched translations need to stay for a short period of time in the prefetch buffer and caches.
We conclude that \chiron{} issues most accurate prefetches within a useful timing window: early enough to hide demand latency, yet not so early that the prefetched translations must remain in the TLB and cache hierarchy for long. 

\begin{figure}[h!]
  \centering
  \includegraphics[width=1.0\columnwidth]{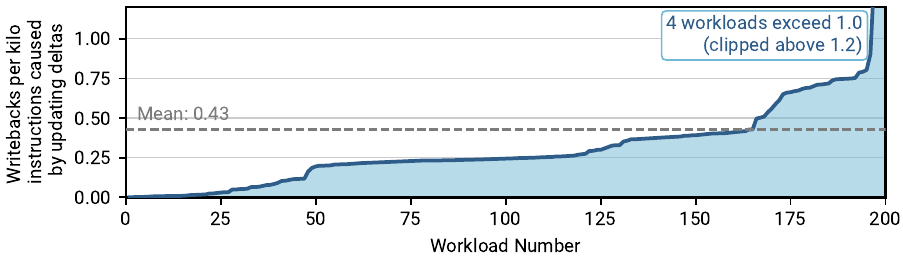}
  \caption{PTE cache block writebacks per kilo instructions caused by \chiron{}'s delta updates, for each of the 200 workloads.}
  \label{fig:pte_writeback_overhead}
\end{figure}

\noindent\textbf{Delta Update Overhead.}
\label{sec:eval:delta_update_overhead}
\chiron{} stores deltas in the leaf PTE cache block. Therefore, updating a delta dirties that block and can cause an extra writeback to DRAM.
To quantify this overhead (which we already incorporate into all our evaluation results), Figure~\ref{fig:pte_writeback_overhead} shows the number of PTE cache block writebacks per kilo instructions (WBPKI) caused by \chiron{}'s delta updates, for each of the 200 workloads, sorted in ascending order.
We observe that the writeback rate is low. On average, \chiron{} causes only 0.43 WBPKI, and 90\% of the workloads experience fewer than 0.7 WBPKI. Only 4 workloads exceed 1.0 WBPKI.
The writeback rate is low for two reasons: (i)~\chiron{} writes the PTE cache block only when the delta changes and (ii)~most deltas stabilize after 10-12 updates (i.e., no further writes are needed for those deltas).
We conclude that \chiron{}'s delta updates cause a small number of writebacks and therefore have small impact on DRAM traffic and overall performance.

\subsection{Multicore Performance}
\label{sec:eval:multicore}

\begin{figure}[h!]
  \centering
  \includegraphics[width=1.0\columnwidth]{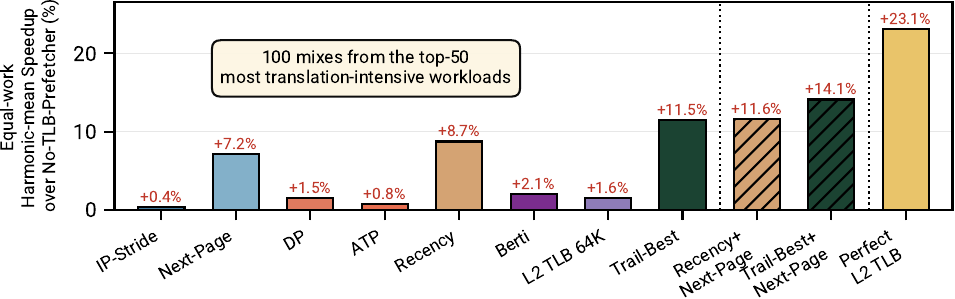}
    \vspace{-3mm}
  \caption{Equal-work harmonic-mean speedup provided by different TLB prefetchers over No-TLB-Prefetcher in a four-core system across 100 multiprogrammed mixes drawn from the 50 most translation-intensive workloads.}
  \label{fig:multicore}
\end{figure}

To evaluate \chiron{} when multiple cores contend for shared resources, we simulate a four-core system running 100 multiprogrammed mixes drawn from the 50 most translation-intensive workloads.
Figure~\ref{fig:multicore} shows the equal-work harmonic-mean speedup~\cite{eeckhout2025use} of each configuration over No-TLB-Prefetcher.
Besides the TLB prefetchers, we evaluate (i) \chiron{} and Recency (the best prior prefetcher) combined with Next-Page, and 
(ii) a system with a much larger L2~TLB (64K entries, $32\times$ No-TLB-Prefetcher's 2048 entries).
We make four observations.

First, \chiron{}-Best improves performance by 11.5\% on average, 2.5\% more than Recency, over the No-TLB-Prefetcher. 
Second, \chiron{}-Best outperforms an L2~TLB with $32\times$ as many entries that provides only a 1.6\% speedup.
Third, \chiron{}-Best closes nearly half of the performance gap to a perfect L2~TLB: its 11.5\% performance improvement captures  50\% of the perfect L2~TLB's improvement (23.1\%).
Fourth, adding Next-Page increases \chiron{}-Best's performance improvement to 14.1\%, 2.2\% more than Recency+Next-Page over the No-TLB-Prefetcher and \chiron{}-Best+Next-Page captures 61\% of the perfect L2~TLB's performance improvement.
We conclude that \chiron{} provides large performance benefits in a multicore system, where shared-resource contention makes address translation more expensive compared to a single-core system.

\subsection{Sensitivity to System Configuration}
\label{sec:eval:system_sensitivity}

We study how three system parameters affect \chiron{}'s benefit: (a)~the fraction of memory backed by large pages, (b)~DRAM bandwidth, and (c)~the number of PTE bits available for \chiron{}'s (and \chiron{}-External) deltas.

\begin{figure}[h!]
  \centering
  \includegraphics[width=0.92\columnwidth]{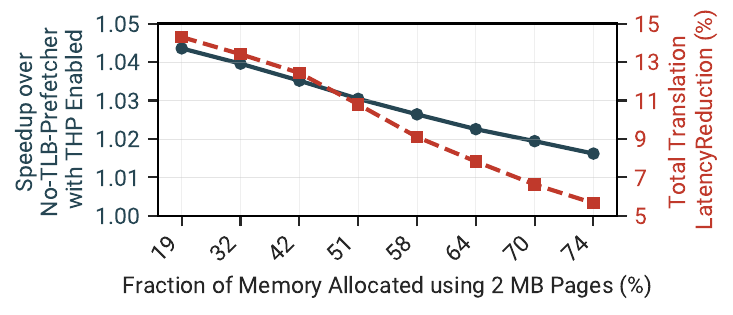}
  \caption{Average speedup (left axis) and total translation latency
    reduction (right axis) provided by \chiron{} over No-TLB-Prefetcher as the fraction of memory backed
    by 2\,MB pages increases, across all 200 workloads, in a system with a 2MB LLC slice per core.}
  \label{fig:frag_sweep}
\end{figure}

\label{sec:eval:large_pages}
\noindent\textbf{(a) Large pages.}
Large pages reduce and accelerate page table walks, and thus the performance headroom for TLB prefetching.
To evaluate how much benefit \chiron{} provides as large pages become more common, we vary memory fragmentation so that the OS backs 19\% to 74\% of memory with 2\,MB pages.
Figure~\ref{fig:frag_sweep} shows the speedup (left axis) and reduction in total translation latency (right axis) provided by \chiron{} over No-TLB-Prefetcher with transparent huge pages (THP~\cite{corbet2011}) enabled and a varying fraction of memory backed by 2\,MB pages, across all 200 workloads, in a system with a 2MB LLC slice per core.
We make two observations.
First, \chiron{}'s benefit decreases as more memory is backed by 2\,MB pages: speedup drops from 4.4\% to 1.6\%, and translation latency reduction drops from 14.3\% to 5.7\%.
The reason is that each 2\,MB TLB entry covers $512\times$ more memory than a 4\,KB entry, which leaves fewer page table walks for \chiron{} to eliminate.
Second, \chiron{} still improves performance by 1.6\% even when 74\% of memory is backed by 2\,MB pages, because the remaining 4\,KB pages still cause expensive page table walks.
We conclude that \chiron{} complements large pages even when a significant fraction of memory is backed by 2\,MB pages.

\begin{figure}[h!]
  \centering
  \includegraphics[width=\columnwidth]{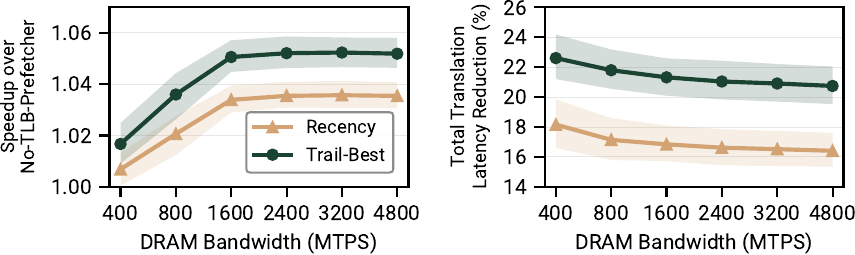}
    \vspace{-3mm}
  \caption{Speedup (left panel) and total translation latency
    reduction (right panel) provided by \chiron{}-Best and Recency over No-TLB-Prefetcher as DRAM bandwidth varies from 400 to 4800\,MTPS, across 200 single-core workloads, in a system with a 2MB LLC slice per core.}
  \label{fig:bw_sweep}
\end{figure}

\label{sec:eval:bandwidth}
\noindent\textbf{(b) DRAM bandwidth.}
\chiron{} issues additional memory requests, so its benefit can depend on the available DRAM bandwidth.
To evaluate this dependence, we vary DRAM bandwidth from 400\,MTPS to 4800\,MTPS.
Figure~\ref{fig:bw_sweep} compares the improvement in performance (left panel) and reduction in total translation latency (right panel) provided by \chiron{}-Best and Recency over No-TLB-Prefetcher, across all 200 workloads (the shaded regions represent the range between the minimum and maximum observed values) in a system with a 2MB LLC slice per core.
We make three observations.
First, at the lowest bandwidth (400\,MTPS), \chiron{} provides its largest translation latency reduction (22.6\%) but its smallest speedup (1.7\%).
The reason is that DRAM contention makes every memory access expensive, address translation is a small fraction of total stall time, and \chiron{}'s prefetches compete with data accesses for scarce bandwidth.
Second, as bandwidth increases, data accesses become cheaper, address translation becomes a larger fraction of stall time, and \chiron{}'s speedup increases, saturating at 5.2\% at 1600\,MTPS and above.
Third, \chiron{}-Best provides higher performance than Recency at every evaluated DRAM bandwidth level, from 1.7\% versus 0.7\% at 400\,MTPS to 5.2\% versus 3.5\% at 1600\,MTPS and above.
We conclude that \chiron{} improves performance across the full range of DRAM bandwidths, from bandwidth-constrained systems to high-bandwidth servers, and consistently outperforms Recency throughout this range.

\label{sec:eval:pte_delta_budget}
\label{sec:eval:separate_metadata}

\noindent\textbf{(c) PTE delta budget and \chiron{}-External.}
To evaluate \chiron{} with fewer PTE bits, we vary the payload of in-PTE \chiron{} from 2 to 10 bits per PTE (Table~\ref{tab:pte_payload_budget}).
To evaluate \chiron{} with no PTE bits, we vary the payload of \chiron{}-External, which stores its payload in the external delta table (\S\ref{sec:mech:external}), from 20 to 176 bits (Table~\ref{tab:trail_external_payload}).
Both tables report the average speedup provided by \chiron{} over No-TLB-Prefetcher across all 200 workloads, in a system with an 8\,MB LLC slice per core.
We make three observations.

First, \chiron{}'s speedup degrades gracefully as the PTE bit budget decreases.
With 18-bit deltas, performance improvement drops from 5.15\% with 4 deltas to 4.84\%, 4.34\%, and 3.47\% with 3, 2, and 1 deltas, respectively.
However, even with only 2-3 bits per PTE, \chiron{} still improves performance.
Second, the number of deltas matters more than the width of each delta.
Widening deltas from 14 to 18 bits increases speedup by at most 0.18 percentage points, whereas adding a new delta increases speedup by up to 0.91 percentage points.
The reason is that most useful deltas fit in 14 bits, whereas each additional delta lets \chiron{} track one more distinct source-to-destination region transition.
Third, \chiron{}-External (Table~\ref{tab:trail_external_payload}) outperforms the best in-PTE configuration when it has more payload bits.
With 4-bit confidence counters and at least 6 slots (132 bits), \chiron{}-External outperforms the 80-bit in-PTE design (5.15\%) and provides up to 5.26\% performance improvement over No-TLB-Prefetcher.
We conclude that \chiron{} does not rely on a specific PTE bit budget and provides significant performance benefit even when all of \chiron{}'s deltas are stored outside the page table.

\begin{table}[h!]
  \centering
  \caption{Speedup provided by in-PTE \chiron{} over No-TLB-Prefetcher, across 200 translation-intensive workloads with an 8\,MB LLC slice per core, as the number of deltas and the width of each delta vary.}
  \label{tab:pte_payload_budget}
  \footnotesize
  \setlength{\tabcolsep}{2pt}
  \renewcommand{\arraystretch}{0.92}
  \begin{tabular*}{\columnwidth}{@{\extracolsep{\fill}}lrrrrr@{}}
    \toprule
    \textbf{Deltas} & \multicolumn{5}{c}{\textbf{$\Delta$ bits}} \\
    \cmidrule(lr){2-6}
    & \textbf{14} & \textbf{15} & \textbf{16} & \textbf{17} & \textbf{18} \\
    \midrule
    1 & +3.29\% & +3.39\% & +3.44\% & +3.46\% & +3.47\% \\
    2 & +4.20\% & +4.27\% & +4.32\% & +4.34\% & +4.34\% \\
    3 & +4.72\% & +4.78\% & +4.82\% & +4.84\% & +4.84\% \\
    4 & +5.03\% & +5.10\% & +5.13\% & +5.15\% & \textbf{+5.15\%} \\
    \bottomrule
  \end{tabular*}
\end{table}

\begin{table}[h!]
  \centering
  \caption{Speedup provided by \chiron{}-External over No-TLB-Prefetcher, across 200 translation-intensive workloads with an 8\,MB LLC slice per core, as the number of deltas and the confidence counter width vary.}
  \label{tab:trail_external_payload}
  \footnotesize
  \setlength{\tabcolsep}{0pt}
  \renewcommand{\arraystretch}{0.88}
  \begin{tabular*}{\columnwidth}{@{\extracolsep{\fill}}rrrrrrr@{}}
    \toprule
    & \multicolumn{2}{c}{\textbf{2-bit conf.}} & \multicolumn{2}{c}{\textbf{3-bit conf.}} & \multicolumn{2}{c}{\textbf{4-bit conf.}} \\
    \cmidrule(lr){2-3}\cmidrule(lr){4-5}\cmidrule(lr){6-7}
    \textbf{Deltas} & \textbf{Bits} & \textbf{Speedup} & \textbf{Bits} & \textbf{Speedup} & \textbf{Bits} & \textbf{Speedup} \\
    \midrule
    1 & 20 & +3.43\% & 21 & +3.41\% & 22 & +3.42\% \\
    2 & 40 & +4.11\% & 42 & +4.19\% & 44 & +4.20\% \\
    3 & 60 & +4.05\% & 63 & +4.59\% & 66 & +4.66\% \\
    4 & 80 & +4.03\% & 84 & +4.77\% & 88 & +4.95\% \\
    5 & 100 & +4.03\% & 105 & +4.71\% & 110 & +5.11\% \\
    6 & 120 & +4.01\% & 126 & +4.67\% & 132 & +5.20\% \\
    7 & 140 & +3.99\% & 147 & +4.62\% & 154 & \textbf{+5.26\%} \\
    8 & 160 & +3.99\% & 168 & +4.62\% & 176 & +5.25\% \\
    \bottomrule
  \end{tabular*}
\end{table}

\subsection{Contribution of \chiron{}'s Components}
\label{sec:eval:design}

\label{sec:eval:ablation}
\begin{figure}[h!]
  \centering
  \includegraphics[width=1.0\columnwidth]{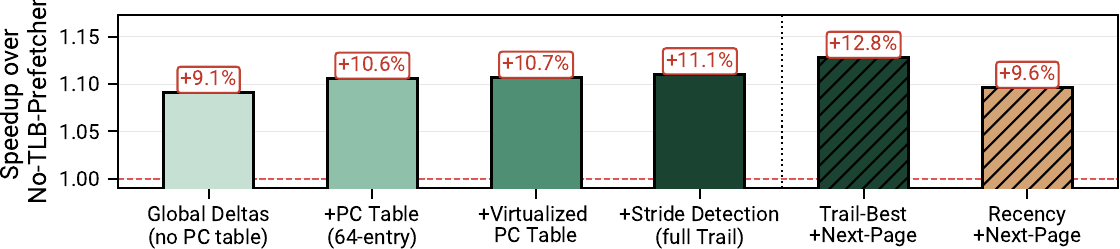}
  \caption{Speedup over the No-TLB-Prefetcher baseline across the 50 most translation-intensive workloads. The first four bars enable \chiron{}'s components one at a time, from global deltas to the full design. The rightmost two bars compare \chiron{}-Best+Next-Page with Recency+Next-Page.}
  \label{fig:ablation}
\end{figure}

To quantify the contribution of each of \chiron{}'s components, we start from a design that learns global deltas without a PC table and enable one component at a time.
Figure~\ref{fig:ablation} shows the performance improvement of each design over the No-TLB-Prefetcher baseline across the 50 most translation-intensive workloads (to highlight the effect of each \chiron{} component in the most demanding applications).
We make five observations.
First, global deltas alone provide a 9.1\% average speedup over No-TLB-Prefetcher across the 50 most translation-intensive workloads.
Second, adding a 64-entry PC table increases speedup to 10.6\%, which shows that per-PC delta tracking improves prediction quality.
Third, virtualizing the PC table preserves learned state across evictions of entries from the PC-table (\S\ref{sec:mech:virt_pc}) and further increases speedup to 10.7\%.
Fourth, virtualization also unlocks stride detection: by preserving per-PC state across evictions, it enables \chiron{} to capture regular translation patterns and improves speedup from 10.7\% to 11.1\%. Even without stride detection, \chiron{} retains 96\% of the full design's performance improvement.
Fifth, Next-Page further complements the full design: \chiron{}-Best+Next-Page improves performance by 12.8\%, 1.7\% more than \chiron{}-Best and 3.2\% more than Recency+Next-Page.
We conclude that region-level temporal prediction provides the foundation of \chiron{}'s benefit. Per-PC tracking and virtualization sharpen that predictor, virtualization unlocks additional gain from stride detection, and Next-Page further complements the full design to deliver the highest overall performance across these translation-intensive workloads.

\section{Related Work}
\label{sec:related}

To our knowledge, this is the first work to: (i)~find that, despite the irregularity of the applications' memory accesses, address translation patterns exhibit a predictable structure (via a large-scale characterization study of 200 modern workloads) and (ii)~propose a new scalable and low-cost TLB prefetcher, \chiron{}, that embeds translation metadata directly in the unused bits of the last-level entries of the page table (PTEs).
We discuss and quantitatively compare \chiron{} against five existing TLB prefetchers~\cite{chen1995tc,Bala1994SoftwarePA,kandiraju2002going,saulsbury2000recency,vavouliotis2021} and a data prefetcher~\cite{navarro2022berti} adapted for TLB prefetching in \S\ref{sec:bg:limitations} and \S\ref{sec:results}, respectively.
In this section, we qualitatively compare \chiron{} to broader classes of address translation optimization techniques.

\vspace{1pt}
\textbf{TLB Management and Hierarchy.}
A broad class of works reduces the frequency of PTWs by increasing translation reach.
Larger or smarter TLB organizations~\cite{papadopoulou2015,latr,juan97,onlinesuperpagepromotionISCA1995,compilerdtlbISPASS2004,vm39,wood1986,skewedTLB,fan05}, additional TLB levels~\cite{sharedl3tlbISCA2011,distlltlbMICRO2018,gpustealing}, and software-managed TLBs~\cite{pomtlbISCA2017,csaltMICRO2017,Bala1994SoftwarePA,softwareTLBNAS2013,softwareTLBISCA2013,bruceMMU1998,softcontrolcachesISCA1986,Nagle1993DesignTF} increase TLB reach.
Replacement policies~\cite{deadTLBHPCA2021,chirpMICRO2020} and cache-hierarchy-backed TLB storage~\cite{kanellopoulosMICRO2023victima,ducati,gputlbreachMICRO2021} keep useful translations as long as possible in the TLB or the cache hierarchy.
These approaches are orthogonal to \chiron{}: they improve how translations are stored and looked up, while \chiron{} predicts \emph{which} translations will be needed next.

\vspace{1pt}
\textbf{Speculative Address Translation.}
Several works accelerate address translation by speculatively predicting the physical addresses corresponding to virtual addresses~\cite{spectlbISCA2011,chloe2020,margaritov2019prefetched,kanellopoulos2026revelator,spec-gpuMICRO2024}.
SpecTLB~\cite{spectlbISCA2011} relies on large page availability of physical memory to accurately predict virtual-to-physical mappings.
SpOT~\cite{chloe2020} exploits contiguous physical allocations to improve the accuracy of speculative address translation.
ASAP~\cite{margaritov2019prefetched} allocates page tables as contiguously as possible to enable page-table-address speculation and overlap accesses to deeper page-table levels within a single walk.
Revelator~\cite{kanellopoulos2026revelator} employs hash-based allocation to enable highly accurate speculative address translation.
These techniques are largely orthogonal to \chiron{}: they accelerate individual page table walks via speculation, whereas \chiron{} predicts future demand translations. Most of them also rely on specialized physical-memory or page-table-layout support to make speculation effective, which \chiron{} does not require.

\vspace{1pt}
\textbf{Large Pages and Contiguity.}
Large pages~\VMlargepages{} and contiguity-based schemes~\VMcontiguity{} reduce the frequency of PTWs by extending TLB reach.
For example, DMT~\cite{dmtASPLOS2024} allocates page tables contiguously and establishes a direct mapping from virtual addresses to last-level PT entries, bypassing intermediate page-table levels, in virtualized environments.
The effectiveness of such approaches depends on the OS's ability to allocate and maintain large contiguous physical regions, which degrades under memory fragmentation~\cite{mansi2024characterizingphysicalmemoryfragmentation,contiguitas2023}.
\chiron{} is orthogonal: it targets the reduction of PTWs via prefetching translations and can be effectively combined with large-page support (\S\ref{sec:eval:large_pages}).

\vspace{1pt}
\textbf{Alternative Page Table Designs.}
Various works~\cite{elastic-cuckoo-asplos20,nestedecht,mehtJovanHPCA2023,hash_dont_cache,distributedptMICRO24osang,kanellopoulos2023utopia,flataAsplos2022,compendiaISMM2021,mitosis-asplos20,chai2025emt,zhao2025learningwalk,vbi} propose new page table designs to reduce the latency of page table walks.
Flattened Page Table~\cite{flataAsplos2022,compendiaISMM2021} and Hash-based Page Table~\cite{elastic-cuckoo-asplos20,nestedecht,mehtJovanHPCA2023,hash_dont_cache,distributedptMICRO24osang,kanellopoulos2023utopia} designs replace pointer chasing-based page table accesses with parallel lookups.
Mitosis~\cite{mitosis-asplos20} replicates and migrates page tables closer to the core to reduce page table walk latencies.
EMT~\cite{chai2025emt} provides an OS framework that enables commodity operating systems to support diverse translation architectures with negligible overhead.
LVM~\cite{zhao2025learningwalk} uses learned indexes to dynamically adapt address translation to the application's virtual address space, effectively providing single-access translation.
While these approaches reduce PTW overhead, they do \emph{not} eliminate PTWs themselves.
\chiron{} is orthogonal to these designs: \chiron{} reduces the number of PTWs and accelerates the remaining ones by prefetching, whereas alternative page table designs reduce the cost of each walk. Thus, \chiron{} can be seamlessly combined with any of these page table designs.

\vspace{1pt}
\textbf{Repurposing Page-Table Metadata.}
Several works use or repurpose PTE bits or related per-page metadata for access tracking and page classification~\cite{kanellopoulos2023utopia}, page-level classification for coherence~\cite{vips}, DRAM-cache footprint hints~\cite{het3}, compression metadata~\cite{comp2}, memory-safety and isolation metadata~\cite{multitag,donky}, message authentication codes~\cite{saxena2023ptguard}, and prefetch filtering~\cite{kamahori2026themis}.
\chiron{} applies this principle to TLB prefetching, embedding translation metadata in unused and/or available PFN bits.
Unlike prior uses that require OS involvement on each access, \chiron{}'s metadata is read and written entirely by hardware during normal page-table walks.

\vspace{1pt}
\textbf{Data Prefetching.}
A rich body of work targets data cache misses through spatial prefetching~\DataPrefetchSpatial{}, temporal and irregular prefetching~\DataPrefetchTemporal{}, learned prefetching~\DataPrefetchLearned{}, and prefetch management~\DataPrefetchMgmt{}.
Closely related latency-hiding mechanisms include cache-level and off-chip prediction~\cite{lp,hermes}, coordination policies that combine off-chip prediction with prefetching~\cite{athena}, and runahead execution~\cite{ramirez2010efficient,dundas,ramirez2008runahead,van2009mlp,hashemi2016continuous,mutlu2003runahead,mutlu2003runahead2,mutlu2006efficient,mutlu2005address,precise_runahead,hashemi2015filtered,vector_runahead}, which generate prefetch-like memory requests by executing ahead of a stalled core.
Some of these techniques can also indirectly warm translation structures: virtual-address-based L1 data prefetchers and runahead execution can trigger translations before the corresponding demand accesses arrive. \chiron{} is complementary to such schemes since it reduces the translation latency of data prefetch requests (e.g., runahead's speculative memory accesses), making them more timely and more effective.

\section{Conclusion}
\label{sec:conclusion}

We introduce \chiron{}, a scalable and low-cost temporal TLB prefetcher that stores its prefetch metadata directly in unused bits of page table entries (PTEs).
\chiron{} builds on a key takeaway from our study of 200 modern translation-intensive workloads: deltas between virtual memory regions that consecutively trigger page table walks exhibit structure that can be exploited for efficient translation prefetching.
Based on this insight, \chiron{} records deltas in the PTEs of each source region and retrieves them during the page table walk at no additional memory access cost.
As a result, \chiron{}'s metadata capacity scales with the application's memory footprint, and \chiron{} requires only a 64-entry table in the processor.
Our evaluation shows that \chiron{} improves performance by 5.7\% (11.5\%) on average in single-core (four-core) systems, outperforming the best prior TLB prefetcher by 1.7\% (2.5\%), across 200 modern workloads.
\chiron{} provides these benefits because it both avoids the most page table walks and shortens the walks it cannot eliminate, while causing negligible metadata writeback traffic.
\chiron{} improves performance when large pages back most of memory, and when no PTE bits are available, by storing its deltas in an external delta table.
More broadly, we believe that the hardware/system-software co-design principles employed in this work, as exemplified by \chiron{}'s use of the page table to store prefetch metadata, can unlock a broad class of powerful prefetching and microarchitectural techniques that were previously impractical, at low hardware cost and with minimal software intervention.
To this end, we open source \chiron{} as part of the Virtuoso ecosystem~\cite{virtuoso_github,kanellopoulos2025virtuoso} at \href{https://github.com/CMU-SAFARI/Virtuoso/tree/trail-artifact-release}{github.com/CMU-SAFARI/Virtuoso/TRAIL}.

\section*{Acknowledgments}

We thank the anonymous reviewers of MICRO 2026, Aamer Jaleel, Tony Nowatzki, Lieven Eeckhout, Timothy Jones, Rakesh Kumar, and Dimitrios Skarlatos for their feedback.
We thank the SAFARI Research Group members for their constructive feedback and for providing a stimulating intellectual and scholarly environment.
We acknowledge the generous gift funding provided by our industrial partners (especially Google, Huawei, Intel, Microsoft, VMware),
which has been instrumental in enabling the research we have been conducting on memory systems.

\pagebreak

\bibliographystyle{unsrt}
\bibliography{references}

\makeatletter
\@ifundefined{UrlBreaks}{}{\g@addto@macro\UrlBreaks{\do\/\do\_\do\-\do\.}}
\makeatother
\providecommand{\aepath}[1]{\nolinkurl{#1}}
\providecommand{\colorbox}[2]{\fbox{#2}}
\newsavebox{\aebox}
\newenvironment{aecmds}%
  {\par\smallskip\noindent
   \begin{lrbox}{\aebox}%
   \begin{minipage}{\dimexpr\linewidth-2\fboxsep\relax}%
   \small\ttfamily\raggedright}%
  {\end{minipage}\end{lrbox}%
   \colorbox{black!6}{\usebox{\aebox}}\par\smallskip}

\appendix
\section{Artifact Appendix}
\label{sec:artifact}
\sloppy

\subsection{Abstract}

This artifact reproduces \chiron{}'s headline results and the measurements that
motivate it. A self-contained HTML code walkthrough of the mechanism (the
\texttt{TemporalPTEPrefetcher} class) is included at \aepath{docs/trail_walkthrough.html}.
\chiron{} is evaluated against the No-TLB-Prefetcher baseline, six prior TLB
prefetchers (IP-Stride, Next-Page, DP, Recency, ATP, and Berti), and a
Perfect L2 TLB upper bound, using Virtuoso~\cite{kanellopoulos2025virtuoso}. The artifact reproduces: the
single-core comparison against prior prefetchers with a 2\,MB and an 8\,MB LLC slice per core (Figure~\ref{fig:prefetcher_comparison}) and the page-table-walk analysis (Figure~\ref{fig:coverage_decomposition}); the in-PTE and \chiron{}-External payload-budget sweeps
(Tables~\ref{tab:pte_payload_budget} and~\ref{tab:trail_external_payload}); the DRAM-bandwidth sensitivity study (Figure~\ref{fig:bw_sweep}) and the component ablation (Figure~\ref{fig:ablation}); the
four-core multiprogrammed study (Figure~\ref{fig:multicore}); and the
motivation figures (Figures~\ref{fig:motivation_pc_cond}, \ref{fig:motivation_patterns}, and \ref{fig:motivation_granularity}--\ref{fig:motivation_bits}), computed directly from
page-table-walk dumps. Evaluation is entirely trace-driven, so no binary instrumentation or proprietary
tools (Intel Pin/SDE) are needed. The workflow is fully scripted: three
setup steps build and validate the artifact, and a single command per suite
then reproduces its figure or table.

\subsection{Artifact check-list (meta-information)}

\begin{itemize}
  \item \textbf{Program:} Virtuoso (Sniper multicore simulator with the MimicOS OS model)
        and the \chiron{} prefetcher (C++17), plus Python analysis and plotting
        scripts. All included in the repository. We suggest using a Debian/Ubuntu Linux distribution.
  \item \textbf{Compilation:} \texttt{g++} with C++17. The instruction decoder
        (libxed) is bundled and built automatically; the trace-replay build
        needs no Intel Pin/SDE and no libtorch. The \texttt{motivation} suite also
        uses a C++17 tool (\texttt{tlb\_prefetch\_sim}), built automatically.
  \item \textbf{Run-time environment:} Linux (tested on Ubuntu~20.04, 22.04, and~24.04).
        Root/\texttt{sudo} is used only to install system packages.
  \item \textbf{Hardware:} any x86-64 machine. A SLURM cluster is recommended
        for the full sweep but is not required; every step also runs on a
        single machine.
  \item \textbf{Metrics:} instructions-per-cycle (IPC), reported as speedup
        over the No-TLB-Prefetcher baseline. Single-core speedups are geometric means
        over the workload suite; the multicore metric is an equal-work
        harmonic mean across the four cores.
  \item \textbf{Output:} each figure or table as PDF and PNG, plus a Markdown
        file with its numbers (e.g., \aepath{figure11.md}), under
        \aepath{experiments/ae/ae_out/}.
  \item \textbf{Experiments:} simulation suites \texttt{head2mb}, \texttt{head8mb},
        \texttt{table5}, \texttt{table6}, \texttt{multicore}, \texttt{abl2mb}, and
        \texttt{head2mb\_mtps400}--\texttt{head2mb\_mtps4800}, plus the trace-free
        \texttt{motivation} suite, all driven by the provided bash harness. (The suite names \texttt{table5} and \texttt{table6} are historical; they reproduce Tables~\ref{tab:pte_payload_budget} and~\ref{tab:trail_external_payload}.)
  \item \textbf{How much disk space required (approximately)?}
        $\sim$300\,GB ($\sim$250\,GB of traces plus the build and results).
  \item \textbf{How much time is needed to prepare workflow (approximately)?}
        $\sim$30--60\,min: a 2--3\,min build plus the (resumable) trace
        download, which dominates and depends on bandwidth.
  \item \textbf{How much time is needed to complete experiments
        (approximately)?} The complete sweep is $\approx$23{,}000 short
        simulations. On a $\sim$1300-core cluster it finishes within
        $\sim$1\,day, and the longest single suite (\texttt{table6}) takes
        $\sim$10\,h. A single suite is far cheaper, and the \texttt{-{}-icount}
        flag shrinks any suite to a minutes-long end-to-end smoke test.
  \item \textbf{Publicly available?} Yes. The source is on GitHub; the traces
        and page-table-walk dumps are public Hugging~Face datasets that need no
        access token.
  \item \textbf{Code licenses (if publicly available)?} Open source; see the
        \aepath{LICENSE} file in the artifact repository.
  \item \textbf{Data licenses (if publicly available)?} Public Hugging~Face
        datasets (traces and dumps).
  \item \textbf{Workflow automation framework used?} A self-contained bash
        harness that handles setup, per-suite execution, and run-all
        orchestration.
  \item \textbf{Archived (provide DOI)?} The artifact is archived on Zenodo at
        \href{https://doi.org/10.5281/zenodo.21541804}{\nolinkurl{doi.org/10.5281/zenodo.21541804}}
        and is also available on the GitHub branch
        \texttt{trail-artifact-release}; the two public Hugging~Face datasets
        below are also required.
\end{itemize}

\subsection{Description}

\subsubsection{How to access}
The source is the \texttt{trail-artifact-release} branch of
\url{https://github.com/CMU-SAFARI/Virtuoso}; all artifact material lives under
\aepath{experiments/ae/}. The traces and the page-table-walk dumps are two
separate public Hugging~Face datasets. The setup script fetches both
(Installation, step~(3)):
\begin{itemize}
  \item Traces and trace-lists:\\
        \href{https://huggingface.co/datasets/konkanello/trail_traces}{\nolinkurl{hf.co/datasets/konkanello/trail_traces}}
  \item Motivation dumps:\\
        \href{https://huggingface.co/datasets/konkanello/trail_ptw_dumps}{\nolinkurl{hf.co/datasets/konkanello/trail_ptw_dumps}}
\end{itemize}

\subsubsection{Hardware dependencies}
An x86-64 machine. The build and the sanity check need $\sim$8\,GB of RAM. 
Each simulation uses a few GB, so more memory and more cores (or a cluster)
simply allow more jobs to run in parallel. About 300\,GB of free disk is needed
for the traces, the build, and the results. A SLURM cluster is optional but
strongly recommended to complete the full sweep in reasonable time.

\subsubsection{Software dependencies}
Linux with a C++17 toolchain and Python~3.8+. All system packages, together
with the Hugging~Face CLI (for the trace and dump downloads) and matplotlib (for the
figures), are installed on Debian/Ubuntu by the provided
\aepath{lib/install_deps.sh} script (equivalent packages exist on other
distributions). Network access is needed to download the traces and to fetch
the decoder during the build. No license-gated or proprietary tools are
required.

\subsubsection{Benchmarks and data sets}
The workload suite spans datacenter server, mobile SoC, graph-analytics, SPEC CPU,
HPC, and industrial traces (Table~\ref{tab:workloads}). The public trace dataset ships the trace files together with
the trace-lists, which the setup script wires into the repository. The
motivation figures use a separate $\sim$3\,GB dataset of per-workload
page-table-walk dumps.

\subsection{Installation}

Run these three steps once. All commands after step~(1) are issued from the
\aepath{experiments/ae/} directory.

\noindent\textbf{(1) Clone the repository and enter the artifact directory.}
\begin{aecmds}
git clone \textbackslash\\
\ \ -{}-branch trail-artifact-release \textbackslash\\
\ \ \url{https://github.com/CMU-SAFARI/Virtuoso.git}\\
cd Virtuoso/experiments/ae
\end{aecmds}

\noindent\textbf{(2) Install the system dependencies} (uses \texttt{sudo} if you
are not root):
\begin{aecmds}
bash lib/install\_deps.sh
\end{aecmds}

\noindent\textbf{(3) Build the simulator, download the traces, and validate.}
This step is resumable: re-run it to continue after any interruption.
\begin{aecmds}
bash build\_and\_validate.sh -{}-skip-deps
\end{aecmds}
Step~(3) performs a trace-only build (no Pin/SDE/libtorch), downloads the public
trace dataset and the page-table-walk dumps, resolves the trace-lists into \aepath{experiments/vm_tlist/}, and
runs a few short simulations on random traces, printing a \texttt{[PASS]}
line with the trace name and its IPC for each. It then stops. This step alone
confirms that the artifact builds and runs end-to-end.

\subsection{Experiment workflow and expected results}

Each paper result is reproduced by a \emph{suite} of simulations, named in the
first column below and selected on the command line with \texttt{-{}-suites}.
The simulation suites fall into four groups: a direct comparison against
prior prefetchers (\texttt{head2mb} and \texttt{head8mb}, the former including the
page-table-walk analysis of Figure~\ref{fig:coverage_decomposition}); sensitivity studies (the
\texttt{table5}/\texttt{table6} payload budgets and the \texttt{head2mb\_mtps*}
DRAM-bandwidth sweep); the component ablation (\texttt{abl2mb}); and a multicore study (\texttt{multicore}).
The \texttt{motivation} suite is trace-free. Each figure or table is written as PDF and PNG,
together with a Markdown file with its numbers, under
\aepath{experiments/ae/ae_out/}; the motivation figures are written under
\aepath{experiments/ae/motivation/motivation_out/}.

\begin{center}\small
\begin{tabularx}{\columnwidth}{@{}l>{\raggedright\arraybackslash}X@{}}
\toprule
suite & paper result and expected outcome \\
\midrule
\texttt{motivation} & Figures~\ref{fig:motivation_pc_cond}, \ref{fig:motivation_patterns}, and \ref{fig:motivation_granularity}--\ref{fig:motivation_bits}; temporal locality of TLB-miss
  successors, the characterization that motivates \chiron{} \\
\texttt{head8mb} & Figure~\ref{fig:prefetcher_comparison} (bottom, 8\,MB); \chiron{} $\approx{+}5.2\%$
  over No-TLB-Prefetcher, ${+}1.8\%$ over the best prior TLB prefetcher, Recency (Perfect L2 TLB $\approx{+}12.0\%$) \\
\texttt{head2mb} & Figure~\ref{fig:prefetcher_comparison} (top, 2\,MB) and Figure~\ref{fig:coverage_decomposition}; \chiron{} $\approx{+}5.7\%$ over
  No-TLB-Prefetcher, ${+}1.7\%$ over the best prior TLB prefetcher, Recency (Perfect L2 TLB $\approx{+}16.7\%$); \chiron{} reduces total demand PTW latency by $\approx$60\% (Recency 42\%) \\
\texttt{table5} & Table~\ref{tab:pte_payload_budget}; in-PTE \chiron{} improves with the payload budget, up to
      $\approx{+}5.15\%$ over No-TLB-Prefetcher \\
\texttt{table6} & Table~\ref{tab:trail_external_payload}; \chiron{}-External improves with the payload budget, up
      to $\approx{+}5.26\%$ over No-TLB-Prefetcher \\
\texttt{head2mb\_mtps*} & Figure~\ref{fig:bw_sweep}; \chiron{} above Recency at every DRAM speed
  ($\approx{+}1.7\%$ vs.\ ${+}0.7\%$ at 400\,MTPS; $\approx{+}5.2\%$ vs.\ ${+}3.5\%$ from 1600\,MTPS) \\
\texttt{abl2mb} & Figure~\ref{fig:ablation}; top-50 workloads: global deltas $\approx{+}9.1\%$, +PC table
  ${+}10.6\%$, +virtualized PC table ${+}10.7\%$, full \chiron{} ${+}11.1\%$ \\
\texttt{multicore} & Figure~\ref{fig:multicore}; \chiron{} reaches $\approx{+}11.5\%$ harmonic-mean
  speedup, versus $\approx{+}8.7\%$ for the best prior TLB prefetcher, Recency
  (Perfect L2 TLB $\approx{+}23.1\%$) \\
\bottomrule
\end{tabularx}
\end{center}

Exact values vary slightly across machines and, for four-core runs, between runs. Single-core results are geometric
means over the workload suite; the multicore result is an equal-work harmonic
mean across the four cores. In every suite \chiron{} should outperform all prior
prefetchers and move toward the Perfect L2 TLB upper bound.

\noindent\textbf{(4) Reproduce the simulation suites.}
\texttt{ae\_run\_all.sh} launches the suites, tracks them in the background, and
renders every figure and table. Use the workflow for your environment: 
\textbf{(4a)} on a SLURM cluster or \textbf{(4b)} on a single machine.

\smallskip
\noindent\textbf{(4a) On a SLURM cluster (recommended).} Run all suites in
parallel:
\begin{aecmds}
bash ae\_run\_all.sh -{}-mode slurm \textbackslash\\
\ \ [-{}-partitions <p>]\\
bash ae\_run\_all.sh -{}-status\\
bash ae\_run\_all.sh -{}-results
\end{aecmds}

Omit \texttt{-{}-partitions} to use the cluster's default partition.
To run only a subset of suites (not recommended), add, e.g., \texttt{-{}-suites "head8mb multicore"}; add \texttt{-{}-icount 2000000} for a minutes-long end-to-end smoke test. Every launch is resumable and re-submits only jobs without valid statistics.

\smallskip
\noindent\textbf{(4b) On a single machine (no SLURM).} Launch as many suites as
you like (all of them, if you want): a single shared scheduler runs their
simulations through the machine's cores, keeping at most \texttt{-{}-jobs}
(default: cores${-}2$) running at a time, so the machine is never oversubscribed:
\begin{aecmds}
bash ae\_run\_all.sh -{}-mode local \textbackslash\\
\ \ -{}-jobs \$(nproc)\\
bash ae\_run\_all.sh -{}-status\\
bash ae\_run\_all.sh -{}-results
\end{aecmds}

The \texttt{-{}-suites} and \texttt{-{}-icount} options work as in (4a).

\smallskip
\noindent\emph{Running a suite by hand is not recommended.}

(a) \texttt{ae\_run\_all.sh} (4a/4b) already performs the whole flow. Internally it
chains three per-suite scripts, exposed only to recover an individual suite that
failed. 

(b) \texttt{ae\_launch.sh} generates the suite's jobs, submits them to SLURM
(or starts a local pool), and returns immediately. 

(c) \texttt{ae\_watch.sh}
detaches into the background and tracks those jobs, writing a live
\aepath{<suite>.status} file and, once every job is accounted for, an
\aepath{<suite>.DONE} pass/fail report. 

(d) \texttt{ae\_results.sh} then parses the
finished statistics and renders the suite's figure and table. 

(e) To heal a failed
suite, re-run \texttt{ae\_launch.sh} (it re-submits only jobs without valid
statistics), then \texttt{ae\_watch.sh}, then \texttt{ae\_results.sh}.

\smallskip
\noindent\textbf{(5) Motivation figures (Figures~\ref{fig:motivation_pc_cond}, \ref{fig:motivation_patterns}, and \ref{fig:motivation_granularity}--\ref{fig:motivation_bits}).}
These come from the page-table-walk dumps that step~(3) downloads, with no simulation.
The \texttt{motivation} suite is part of step~(4), so \texttt{ae\_run\_all.sh} already
produces them. To run them on their own (analysis, then plotting):
\begin{aecmds}
bash motivation/run\_motivation.sh \textbackslash\\
\ \ -{}-mode slurm [-{}-partitions <p>]\\
\# single machine: -{}-mode local -{}-jobs \$(nproc)\\
bash motivation/run\_motivation.sh -{}-plot
\end{aecmds}
Each workload's analysis is stored as JSON files, so the step is resumable;
\texttt{-{}-plot} renders the figures once all workloads are analyzed.
\subsection{Notes}

\begin{itemize}
  \item Every phase is resumable: re-running any of steps~(3)--(5) skips
        whatever is already complete.
  \item The full artifact instructions are in \aepath{experiments/ae/README.md},
        and the motivation-specific instructions in
        \aepath{experiments/ae/motivation/README.md}.
\end{itemize}

\end{document}